\documentclass[prd,aps,showpacs,nofootinbib,preprint,eqsecnum]{revtex4}
\usepackage{amsmath}
\usepackage{amssymb}
\usepackage{amsfonts}
\usepackage{graphicx,bm}
\usepackage{dcolumn}
\usepackage{color,amsxtra}
\usepackage{epsf}
\usepackage{enumerate}
\usepackage{hhline}
\usepackage{array}
\usepackage{tabularx}
\newcommand{\be}{\begin{equation}}
\newcommand{\ee}{\end{equation}}
\newcommand{\bea}{\begin{eqnarray}}
\newcommand{\eea}{\end{eqnarray}}
\newcommand{\beaa}{\begin{eqnarray*}}
\newcommand{\eeaa}{\end{eqnarray*}}

\newcommand{\nn}{\nonumber \\}
\newcommand{\e}{\mathrm{e}}

\newcommand{\Eqn}[1]{&\hspace{-0.2em}#1\hspace{-0.2em}&}

\def\be{\begin{equation}}
\def\ee{\end{equation}}
\def\bea{\begin{eqnarray}}
\def\eea{\end{eqnarray}}

\def\nn{\nonumber \\}
\def\e{\mathrm{e}}

\begin{document}


\title{Domain wall solution in $F(R)$ gravity and variation of the fine 
structure constant}

\author{Kazuharu Bamba$^{1,}$\footnote{
E-mail address: bamba@kmi.nagoya-u.ac.jp},
Shin'ichi Nojiri$^{1, 2,}$\footnote{E-mail address:
nojiri@phys.nagoya-u.ac.jp} 
and 
Sergei D. Odintsov$^{3,}$\footnote{
E-mail address: odintsov@ieec.uab.es}$^{,}$\footnote{Also at Tomsk 
State Pedagogical University.}}
\affiliation{
$^1$Kobayashi-Maskawa Institute for the Origin of Particles and the
Universe,
Nagoya University, Nagoya 464-8602, Japan\\
$^2$Department of Physics, Nagoya University, Nagoya 464-8602, Japan\\
$^3$Instituci\`{o} Catalana de Recerca i Estudis Avan\c{c}ats (ICREA)
and Institut de Ciencies de l'Espai (IEEC-CSIC),
Campus UAB, Facultat de Ciencies, Torre C5-Par-2a pl, E-08193 Bellaterra
(Barcelona), Spain 
}


\begin{abstract}

We construct a domain wall solution in $F(R)$ gravity. 
We reconstruct a static domain wall solution 
in a scalar field theory. 
We also reconstruct an explicit $F(R)$ gravity model in which 
a static domain wall solution can be realized. 
Moreover, we show that there could exist an effective (gravitational) domain 
wall in the framework of $F(R)$ gravity. 
In addition, it is demonstrated that a logarithmic 
non-minimal gravitational coupling of the electromagnetic theory 
in $F(R)$ gravity may produce time-variation of the fine structure constant 
which may increase with decrease of the curvature, 
and that this model would be ruled out by the constraints on the time variation of the fine structure constant from quasar absorption lines. 
We also present cosmological consequences of the coupling of the 
electromagnetic field to a scalar field as well as the scalar curvature 
and discuss the relation between variation of the fine structure constant and 
the breaking of the conformal invariance of the electromagnetic field. 

\end{abstract}

\pacs{
04.50.Kd, 95.36.+x, 98.80.Cq
}

\maketitle

\section{Introduction}

According to 
recent cosmological observations, e.g., 
Supernovae Ia (SNe Ia)~\cite{SN1}, 
cosmic microwave background (CMB) radiation~\cite{WMAP, Komatsu:2010fb}, 
large scale structure (LSS)~\cite{LSS},  
baryon acoustic oscillations (BAO)~\cite{Eisenstein:2005su}, 
and weak lensing~\cite{Jain:2003tba}, 
it has been implied that the current expansion of the universe 
is accelerating. 
Studies on 
the late time cosmic acceleration are classified into the representative 
two categories. 
One is to introduce dark energy such as cosmological constant 
in the framework of general relativity 
(for a recent review, see, e.g.,~\cite{Li:2011sd}). 
The other is to modify the gravitational theory, 
for instance, $F(R)$ gravity, 
where $F(R)$ is an arbitrary function of the scalar curvature $R$ 
(for recent reviews on $F(R)$ gravity, 
see, e.g.,~\cite{Review-Nojiri-Odintsov, Book-Capozziello-Faraoni, 
Clifton:2011jh}). 

Recently, 
not only temporal~\cite{Time-variation, Murphy:2003hw} 
but also spatial~\cite{Webb:2010hc} variations of 
the fine structure constant ${\alpha}_{\mathrm{EM}}$ 
have been suggested. 
To account for the spatial variation of ${\alpha}_{\mathrm{EM}}$, 
the signature of a domain wall produced in the spontaneous symmetry 
breaking involving a dilaton-like scalar field coupled to 
electromagnetism has been considered in Ref.~\cite{Olive:2010vh}. 
Furthermore, 
in Ref.~\cite{Chiba:2011en} it has been shown that 
a runaway domain wall, 
which is formed by a runaway type potential of 
a scalar field~\cite{Cho:1998jk}, 
can explain both 
the time variation by its potential 
and the spatial one by its formation 
simultaneously. 
It is interesting to note that 
in Ref.~\cite{M-B-C} time and spatial variations of coupling constant 
have been studied and that 
when the chameleon field is introduced, variations of coupling constant is 
related to the chameleon mechanism~\cite{Chameleon-mechanism}. 

On the other hand, 
a domain wall solution in the framework of $F(R)$ gravity has not 
been investigated in detail yet. In particular, it is interesting to 
reconstruct an $F(R)$ gravity model in which a domain wall solution can be 
realized. 
It is known that $F(R)$ gravity can be written as 
a corresponding scalar field theory through 
a conformal transformation to the Einstein frame. 
In this paper, 
we reconstruct an explicit $F(R)$ gravity model in which 
a static domain wall solution can be realized. 
First, by using a procedure proposed in Ref.~\cite{Capozziello:2005tf}, 
we reconstruct a static domain wall solution in a scalar field theory. 
Next, in a similar configuration, 
we reconstruct an explicit form of $F(R)$ with forming 
a static domain wall solution. 
Moreover, by applying the method of reconstruction of $F(R)$ gravity 
in Ref.~\cite{Reconstruction-F(R)-N-O}, we show that 
there could exist an effective (gravitational) domain wall 
in the framework of $F(R)$ gravity. 
In addition, 
we discuss an issue of a connection between $F(R)$ gravity 
and variation of the fine structure constant by exploring 
non-minimal Maxwell-$F(R)$ gravity. 
Furthermore,
we present cosmological consequences of the coupling of the electromagnetic 
field to a scalar field as well as the scalar curvature. 
We also study the relation between 
variation of the fine structure constant and 
the breaking of the conformal invariance of the electromagnetic field. 
We use units of $k_\mathrm{B} = c = \hbar = 1$ and denote the
gravitational constant $8 \pi G$ by
${\kappa}^2 \equiv 8\pi/{M_{\mathrm{Pl}}}^2$
with the Planck mass of $M_{\mathrm{Pl}} = G^{-1/2} = 1.2 \times 10^{19}$GeV. 
Moreover, in terms of electromagnetism we adopt Heaviside-Lorentz units. 

The paper is organized as follows. 
In Sec.\ II, we describe $F(R)$ gravity and a corresponding scalar field 
theory by using a conformal transformation of $F(R)$ gravity 
to the Einstein frame. 
In Sec.\ III, we reconstruct a static domain wall solution 
in a scalar field theory. 
In Sec.\ IV, 
we also reconstruct an explicit $F(R)$ gravity model in which 
a static domain wall solution can be realized. 
In Sec.\ V, we demonstrate that there could exist 
an effective (gravitational) domain wall in $F(R)$ gravity. 
In Sec.\ VI, 
we consider non-minimal Maxwell-$F(R)$ gravity 
and examine a relation between $F(R)$ gravity 
and variation of the fine structure constant. 
In addition, 
we investigate cosmological consequences of the coupling of 
the electromagnetic field to a scalar field as well as the scalar curvature 
in Sec.\ VII. 
Finally, conclusions are given in Sec.\ VIII. 

\section{Comparison of $F(R)$ gravity with 
a scalar field theory having a runaway type potential 
}

\subsection{$F(R)$ gravity and a corresponding scalar field theory}

The action of $F(R)$ gravity with matter is written as 
\begin{equation} 
S = \int d^4 x \sqrt{-g} 
\frac{F(R)}{2\kappa^2} + 
\int d^4 x 
{\mathcal{L}}_{\mathrm{M}} \left( g_{\mu\nu}, {\Psi}_{\mathrm{M}} \right)\,,
\label{eq:2.1}
\end{equation}
where $g$ is the determinant of the metric tensor $g_{\mu\nu}$ and 
${\mathcal{L}}_{\mathrm{M}}$ is the matter Lagrangian. 

We make a conformal transformation to the Einstein frame:
\begin{equation} 
\tilde{g}_{\mu\nu} = \Omega^2 g_{\mu\nu}\,,
\label{eq:2.2}
\end{equation}
where 
\begin{eqnarray} 
\Omega^2 \Eqn{\equiv} F_{,R}\,, 
\label{eq:2.3} \\ 
F_{,R} \Eqn{\equiv} \frac{d F(R)}{d R}\,.
\label{eq:2.4} 
\end{eqnarray}
Here, a tilde represents quantities in the Einstein frame. 
We introduce a new scalar field $\phi$, defined by 
\begin{equation} 
\phi \equiv \sqrt{\frac{3}{2}} \frac{1}{\kappa} \ln F_{,R}\,.
\label{eq:2.5}
\end{equation}
The relation between $R$ and $\tilde{R}$ is expressed as 
\begin{equation} 
R = e^{1/\sqrt{3} \kappa \phi} 
\left[ \tilde{R} + \sqrt{3} \tilde{\Box} \left( \kappa \phi \right) 
- \frac{1}{2} \tilde{g}^{\mu\nu}
{\partial}_{\mu} \left( \kappa \phi \right) 
{\partial}_{\nu} \left( \kappa \phi \right)
\right]\,, 
\label{eq:2-Add-1}
\end{equation}
where 
\begin{equation} 
\tilde{\Box} \left( \kappa \phi \right) 
= \frac{1}{\sqrt{-\tilde{g}}} {\partial}_{\mu}
\left[ \sqrt{-\tilde{g}} \tilde{g}^{\mu\nu} {\partial}_{\nu} 
\left( \kappa \phi \right) \right]\,.
\label{eq:2-Add-2}
\end{equation}
The action in the Einstein frame is given by~\cite{F-M} 
\begin{equation} 
S_{\mathrm{E}} = 
\int d^4 x \sqrt{-\tilde{g}} \left( \frac{\tilde{R}}{2\kappa^2} - 
\frac{1}{2} \tilde{g}^{\mu\nu} {\partial}_{\mu} \phi {\partial}_{\nu} \phi 
- V(\phi) \right) + 
\int d^4 x 
{\mathcal{L}}_{\mathrm{M}} 
\left( \left( F_{,R} \right)^{-1}(\phi) \tilde{g}_{\mu\nu}, 
{\Psi}_{\mathrm{M}} \right)\,,
\label{eq:2.6}
\end{equation}
where 
\begin{equation} 
V(\phi) = \frac{F_{,R}\tilde{R}-F}{2\kappa^2 \left(F_{,R}\right)^2}\,.
\label{eq:2.7}
\end{equation}
%

\subsection{Runaway domain wall and a varying fine structure constant 
${\alpha}_{\mathrm{EM}}$} 

In Ref.~\cite{Chiba:2011en}, 
the following action describing a runaway domain wall 
and a space-time varying fine structure constant 
${\alpha}_{\mathrm{EM}}$ has been proposed: 
%
\begin{eqnarray} 
S_{\mathrm{E}} \Eqn{=} 
\int d^4 x \sqrt{-\tilde{g}} \left( \frac{\tilde{R}}{2\kappa^2} - 
\frac{1}{2} \tilde{g}^{\mu\nu} {\partial}_{\mu} \phi {\partial}_{\nu} \phi 
- V(\phi) \right) +
\int d^4 x \sqrt{-\tilde{g}} 
\left(
-\frac{1}{4} B(\phi) \tilde{g}^{\mu\alpha} \tilde{g}^{\nu\beta}
F_{\mu\nu} F_{\alpha\beta} 
\right) 
\nonumber \\ 
&& 
{}+ S_{\mathrm{matter}}\,,
\label{eq:2.8}
\end{eqnarray}
%
where 
\begin{eqnarray} 
V(\phi) \Eqn{=} \frac{M^{2p+4}}{\left( \phi^2 + \sigma^2 \right)^p}\,, 
\label{eq:2.9} \\  
B(\phi) \Eqn{=} \e^{-\xi \kappa \phi}\,. 
\label{eq:2.10} \\ 
F_{\mu\nu} \Eqn{=} {\partial}_{\mu}A_{\nu} - {\partial}_{\nu}A_{\mu}\,. 
\label{eq:2.11}
\end{eqnarray}
Here, 
$F_{\mu\nu}$ 
is the electromagnetic field-strength tensor and 
$A_{\mu}$ is the $U(1)$ gauge field. 
$S_{\mathrm{matter}}$ is the action for other ordinary matters. 
Moreover, 
$V(\phi)$ is a scalar field potential of runaway type, 
$M$ is a mass scale, 
$p (>1)$ is a constant assumed to be larger than unity, 
$\sigma (< \phi)$ is a constant assumed to be smaller than the 
value of $\phi$. 
It is known that although there is no minima in the potential $V(\phi)$, 
the discrete symmetry $\phi \leftrightarrow -\phi$ can be broken dynamically 
and consequently a domain wall can be formed. 
Furthermore, $B(\phi)$ is a coupling function of $\phi$ to the 
electromagnetic kinetic term 
and $\xi$ is a constant. 
The spatio-temporal variations of ${\alpha}_{\mathrm{EM}}$ 
come from the variation of $B(\phi)$ in terms of 
space and time 
because ${\alpha}_{\mathrm{EM}} (\phi) = 
{\alpha}_{\mathrm{EM}}^{(0)}/B(\phi)$, 
where ${\alpha}_{\mathrm{EM}}^{(0)} = e^2/\left(4\pi \right)$ with 
$e$ being the charge of the electron~\cite{Kolb and Turner}, 
is the bare fine structure constant, 
and $\xi$ is a constant. 
We note that since the electromagnetic fields have the conformal invariance, 
the conformal transformation in Eq.~(\ref{eq:2.2}) does not generate 
the non-trivial coupling of the scalar filed $\phi$ with the electromagnetic 
fields. 

The current value of the Hubble parameter is given by  
$H_0 = 2.1 h \times 10^{-42} \mathrm{GeV}$~\cite{Kolb and Turner}
with $h = 0.7$~\cite{Komatsu:2010fb, Freedman:2000cf}. 
We assume the flat Friedmann-Lema\^{i}tre-Robertson-Walker (FLRW) metric 
\begin{equation}  
ds^2 = - dt^2 + a^2(t) \sum_{i=1,2,3}\left(dx^i\right)^2\,. 
\label{eq:2.18}
\end{equation}
In this background, $R = 6\dot{H} + 12H^2$, where 
$H=\dot{a}/a$ is the Hubble parameter and 
the dot denotes the time derivative of $\partial/\partial t$. 
Hence, the current curvature $R_0$ is $R_0 \approx 12 H_0^2$.

\section{Reconstruction of a static domain wall solution 
in a scalar field theory}

In this section, 
we reconstruct a static domain wall solution in a scalar field theory 
by using a procedure in Ref.~\cite{Capozziello:2005tf}. 

We consider the following action:
\be
\label{I1}
S = \int d^D x \sqrt{-g} \left( \frac{R}{2\kappa^2} - \frac{\omega(\varphi)}{2}\partial_\mu \varphi \partial^\mu \varphi 
 - \mathcal{V} (\varphi) \right)\, .
\ee
We also assume the following $D=d+1$ dimensional warped metric
\be
\label{I2}
ds^2 = dy^2 + \e^{u(y)} \sum_{\mu,\nu=0}^{d-1} {\hat g}_{\mu\nu} dx^\mu dx^\nu\, ,
\ee
and we also assume the scalar field only depends on $y$. 
In (\ref{I2}), ${\hat g}_{\mu\nu}$ is the metric of the $d$-dimensional Einstein manifold defined by 
${\hat R}_{\mu\nu} = \frac{d-1}{l^2} {\hat g}_{\mu\nu}$. The de Sitter space corresponds to $1/l^2>0$, 
the anti-de Sitter space to $1/l^2<0$, and the flat space $1/l^2 = 0$. 
Then the $(y,y)$ component and $(\mu,\nu)$ component of the Einstein equation are given by
\bea
\label{I3}
 - \frac{d(d-1)}{2l^2} \e^{-u} + \frac{d(d-1)}{8} \left( u'\right)^2 
 &=& \frac{1}{2}\omega(\varphi) \left(\varphi'\right)^2 - \mathcal{V}(\varphi) \, , \\
\label{I4}
 - \frac{(d-1)(d-2)}{2l^2} \e^{-u} + \frac{d-1}{2} u'' + \frac{d(d-1)}{8} \left( u'\right)^2 
 &=& - \frac{1}{2}\omega(\varphi) \left(\varphi'\right)^2 - \mathcal{V}(\varphi) \, , 
\eea
where the prime denotes the derivative with respect to $y$. 
Now we may choose $\varphi=y$\footnote{
The reason why we may choose $\varphi=y$ is as follows. 
We here examine the case in which the scalar field $\varphi$ only depends on 
$y$. As a simplest choice, we take $\varphi=y$. 
Even if we choose other form such as $\varphi = \varphi (y)$, 
by using a variable transformation, we can rewrite the action to 
the one represented with $\varphi=y$. 
Hence, all the consequences, e.g., when $y$ goes to infinity, 
$u$ becomes zero, would not be depend on the choice of the form of $\varphi$ 
qualitatively. 
}. 
In this case, we also take $\kappa^2 = 1$. 
Then Eqs.~(\ref{I3}) and (\ref{I4}) give 
\bea
\label{I5}
\omega(\varphi) &=& - \frac{d-1}{2}u'' - \frac{d-1}{l^2} \e^{-u}\, , \\
\label{I6}
\mathcal{V}(\varphi) &=& - \frac{d-1}{4} u'' - \frac{d(d-1)}{8} \left( u' \right)^2 + \frac{(d-1)^2}{2l^2}\e^{-u}\, .
\eea
The energy density $\rho$ is now given by
\be
\label{I7}
\rho = \frac{\omega(\varphi)}{2}\left( \varphi' \right)^2 + \mathcal{V}(\varphi)
= - \frac{d-1}{2} u'' - \frac{d(d-1)}{8} \left( u' \right)^2 + \frac{(d-1)(d-2)}{2l^2}\e^{-u}\, .
\ee
When we assume the $D$ dimensional space is flat, we find $u\to 0$ when $\left| y \right| \to \infty$, 
the second term dominates in (\ref{I5}), $\omega(\varphi) \sim - (d-1) /l^2$. When $\omega(\varphi)$ is negative, 
which corresponds to $1/l^2>0$, the scalar field $\varphi$ becomes a ghost. 
In case of $1/l^2=0$, we find $\omega(\varphi) = - (d-1)u''/2$. Then if we assume $Z_2$ symmetry 
of the metric, which is the invariance under the transformation $y\to -y$, there must be a region where 
$\omega(\varphi)$ becomes negative and therefore $\varphi$ becomes a ghost. 
We should 
note that the energy density often becomes negative. 
Anyway if we admit the ghost and negative energy density, for arbitrary $u$, we find a model which admits that $u$ 
as a solution of the Einstein equation. 
For example, we may consider
\be
\label{I8}
u = u_0 \e^{-y^2/y_0^2}\, ,
\ee
with constants $u_0$ and $y_0$. Then if we consider the model
\bea
\label{I9}
\omega(\varphi) &=& - (d-1)\left( \frac{2\varphi^2}{y_0^4} - \frac{1}{y_0^2} \right)\e^{- \varphi^2/y_0^2} 
 - \frac{(d-1)}{l^2}\e^{-u_0 \e^{-\varphi^2/y_0^2}}\, ,\nn
\mathcal{V}(\varphi) &=& - \frac{d-1}{2}\left( \frac{2\varphi^2}{y_0^4} - \frac{1}{y_0^2} \right)\e^{- \varphi^2/y_0^2} 
+ \frac{(d-1)^2}{l^2}\e^{-u_0 \e^{-\varphi^2/y_0^2}}\, ,
\eea
we obtain $u$ in (\ref{I8}) as a solution of the Einstein equation. For the model, the distribution of the 
energy density is given by
\be
\label{I10}
\rho(y) = - \frac{d-1}{2}\left( \frac{2y^2}{y_0^4} - \frac{1}{y_0^2} \right)\e^{- y^2/y_0^2} 
+ \frac{(d-1)^2}{l^2}\e^{-u_0 \e^{-y^2/y_0^2}}\, ,
\ee
which is localized at $y\sim 0$ and makes a domain wall. 

As a consequence, 
we have reconstructed the forms of $\omega(\varphi)$ and $\mathcal{V}(\varphi)$ in (\ref{I9}) so that the metric in Eq.~(\ref{I2}) with Eq.~(\ref{I8}) 
can be a solution of the Einstein equation. 
In this model, we have illustrated that the distribution of the 
energy density given by Eq.~(\ref{I10}) is localized at $y\sim 0$ and hence 
it can be regarded as a domain wall. 
A condition for the localization of the energy density is that 
in the limit of $\left| y \right| \to \infty$, the asymptotic behavior of 
$u\to 0$ is satisfied. 

In this work, as a first step to demonstrate whether 
the configuration that 1-dimensional domain wall and 
$d$-dimensional Einstein manifold (e.g., for an ordinary 4-dimensional space-time, $d=3$) can be realized in the framework of a scalar field theory, 
we explore a static domain wall, provided a $D=d+1$ dimensional warped metric 
in Eq.~(\ref{I2}). 
In order to analyze the stability of these domain walls against small 
perturbations, i.e., the time evolution of the above configuration, 
it is necessary to consider a different metric including a time component from 
that in Eq.~(\ref{I2}). 
In this work, the existence of domain wall solutions (as an assumption) is 
only studied, while leaving the stability to future work. 

Furthermore, 
we should caution that since there exist the situation in which the scalar 
field $\varphi$ becomes a ghost, in the sense of quantum theory 
a static domain wall solution in a scalar field theory reconstructed in 
this section is not physically viable. 
Nevertheless, the motivation to carry on and make the subsequent analysis on 
this model is as follows. 
It would be important to explore whether 
the distribution of the energy density is localized so that 
such a configuration could be a domain wall, which might correspond to 
so-called a brane in the literature. 

\section{Reconstruction of an explicit $F(R)$ gravity model realizing 
a static domain wall solution}

In this section, in a similar configuration to that in Sec.~III,  
we reconstruct an explicit $F(R)$ gravity model realizing 
a static domain wall solution.

\subsection{Gravitational field equations}

Varying the action in Eq.~(\ref{eq:2.1}) with respect to $g_{\alpha \beta}$, 
we obtain 
\begin{equation}
-\frac{1}{2} F g_{\alpha \beta} 
+ \left( 
R_{\alpha \beta} - \nabla_{\alpha}\nabla_{\beta} + g_{\alpha \beta} \Box 
\right) F_{,R} 
= 
\kappa^2 T^{(\mathrm{M})}_{\alpha \beta}\,. 
\label{eq:4.1}
\end{equation}
where 
${\nabla}_{\alpha}$ is the covariant derivative operator associated with 
$g_{\alpha \beta}$, 
$\Box \equiv g^{\alpha \beta} {\nabla}_{\alpha} {\nabla}_{\beta}$
is the covariant d'Alembertian for a scalar field, 
and 
$T^{(\mathrm{M})}_{\alpha \beta} 
\equiv 
-\left(2/\sqrt{-g}\right)
\left( \delta 
{\mathcal{L}}_{\mathrm{M}} /\delta g^{\alpha \beta} 
\right)$ is the energy-momentum tensor of matter 
and given by 
$
{T^\alpha}_{\beta}^{(\mathrm{M})} 
= \mathrm{diag} 
(-{\rho}_{\mathrm{M}}, {P}_{\mathrm{M}}, {P}_{\mathrm{M}}, {P}_{\mathrm{M}})$ 
with ${\rho}_{\mathrm{M}}$ and ${P}_{\mathrm{M}}$ being 
the energy density and pressure of matter, respectively. 

Equation (\ref{eq:4.1}) can be described as 
\begin{equation}
G_{\alpha \beta} = \kappa^2 \left( T^{(\mathrm{M})}_{\alpha \beta} 
+ T^{(\mathrm{D})}_{\alpha \beta} \right)\,, 
\label{eq:4.2}
\end{equation}
where 
\begin{equation}
\kappa^2 T^{(\mathrm{D})}_{\alpha \beta} 
\equiv 
\frac{1}{2}\left(F-R\right) 
g_{\alpha \beta} + \left( 1 - F_{,R} \right) R_{\alpha \beta} 
+ \left( \nabla_{\alpha}\nabla_{\beta} - g_{\alpha \beta} \Box 
\right) F_{,R}\,. 
\label{eq:4.3}
\end{equation}
Here, $G_{\alpha \beta} \equiv R_{\alpha \beta} -\left(1/2\right) 
g_{\alpha \beta} R$ is the Einstein tensor 
and 
$\kappa^2 T^{(\mathrm{D})}_{\alpha \beta}$ 
can be regarded as the contribution to the energy-momentum tensor 
from the deviation of $F(R)$ gravity from general relativity. 

We take the $D=d+1$ dimensional warped metric in Eq.~(\ref{I2}), 
in which $g_{yy} = 1$ and $g_{\mu\nu} = \e^u \hat{g}_{\mu\nu}$. 
The $(y,y)$ component and the trace of $(\mu,\nu)$ component of 
the gravitational field equation (\ref{eq:4.1}) 
are given by 
%
\begin{eqnarray} 
&&
\frac{d-1}{2} u' \left( F_{,R} \right)' 
-\frac{d}{2} \left[ u'' + \frac{1}{2}\left(u'\right)^2 
\right] F_{,R} -\frac{1}{2} F = \kappa^2 T^{(\mathrm{M})}_{yy}\,, 
\label{eq:4.4} \\ 
%
%
&&
d \left( F_{,R} \right)'' 
+ \frac{d\left(d-2\right)}{2} u' 
\left( F_{,R} \right)' 
+ \left\{ 
-\frac{d}{2} \left[ u'' 
+ \frac{d}{2}\left(u'\right)^2 \right] 
+ \frac{d\left(d-1\right)}{l^2} \e^{-u}
\right\} F_{,R} -\frac{d}{2} F 
\nonumber \\ 
&&
{}= \kappa^2 
\sum_{\mu,\nu=0}^{d-1} 
g^{\mu\nu} T^{(\mathrm{M})}_{\mu\nu}\,. 
\label{eq:4.5}
\end{eqnarray}
%
where $\left( F_{,R} \right)' \equiv d F_{,R}/d y$ 
and $\left( F_{,R} \right)'' \equiv d^2 F_{,R}/d y^2$. 

Moreover, in the background described by Eq.~(\ref{I2}), 
$R$ is expressed as 
\begin{equation} 
R = -\frac{d}{2} \left[ 2u'' 
+ \frac{1+d}{2}\left(u'\right)^2 \right] 
+ \frac{d\left(d-1\right)}{l^2} \e^{-u}\,. 
\label{eq:4.6}
\end{equation}

We rewrite Eqs.~(\ref{eq:4.4}) and (\ref{eq:4.5}) in the 
form of Eq.~(\ref{eq:4.2}) as follows: 
\begin{eqnarray} 
\hspace{-5mm}
&&
-\frac{d}{2} \left[ u'' + \frac{1}{2}\left(u'\right)^2 
\right] -\frac{R}{2} = \kappa^2 \left( T^{(\mathrm{M})}_{yy} 
+ T^{(\mathrm{D})}_{yy} \right)\,, 
\label{eq:4.7} \\
\hspace{-5mm}
&&
-\frac{d}{2} \left[ u'' 
+ \frac{d}{2}\left(u'\right)^2 \right] 
+ \frac{d\left(d-1\right)}{l^2} \e^{-u}
-\frac{d}{2} R 
= \kappa^2 \left( 
\sum_{\mu,\nu=0}^{d-1} 
g^{\mu\nu} T^{(\mathrm{M})}_{\mu\nu} 
+ \sum_{\mu,\nu=0}^{d-1} g^{\mu\nu} T^{(\mathrm{D})}_{\mu\nu} 
\right)\,, 
\label{eq:4.8}
\end{eqnarray}
where 
\begin{eqnarray} 
\hspace{-5mm}
\kappa^2 T^{(\mathrm{D})}_{yy} \Eqn{\equiv} 
-\frac{d-1}{2} u' \left( F_{,R} \right)' 
+\frac{d}{2} \left[ u'' + \frac{1}{2}\left(u'\right)^2 
\right] \left( F_{,R} - 1 \right) +\frac{1}{2} \left(F-R\right)\,, 
\label{eq:4.9} \\ 
%
%
\hspace{-5mm}
\kappa^2 \sum_{\mu,\nu=0}^{d-1} g^{\mu\nu} T^{(\mathrm{D})}_{\mu\nu} 
\Eqn{\equiv} 
-d \left( F_{,R} \right)'' 
-\frac{d\left(d-2\right)}{2} u' 
\left( F_{,R} \right)' 
\nonumber \\ 
\hspace{-5mm}
&&
{}+ \left\{ 
\frac{d}{2} \left[ u'' 
+ \frac{d}{2}\left(u'\right)^2 \right] 
- \frac{d\left(d-1\right)}{l^2} \e^{-u}
\right\} \left( F_{,R} -1 \right) +\frac{d}{2} \left(F-R\right)\,. 
\label{eq:4.10}
\end{eqnarray}
By substituting Eq.~(\ref{eq:4.6}) into 
Eqs.~(\ref{eq:4.7}) and (\ref{eq:4.8}), 
we see that the left-hand side (l.h.s.) of Eq.~(\ref{eq:4.7}) 
is equal to that of Eq.~(\ref{I3}) and 
the l.h.s. of Eq.~(\ref{eq:4.8}) divided by $d$ 
is equal to that of Eq.~(\ref{I4}). 
We note that Eqs.~(\ref{eq:4.7}) and (\ref{eq:4.8}) are exactly equivalent to 
Eqs.~(\ref{eq:4.4}) and (\ref{eq:4.5}), respectively. 
By comparing these equations with Eqs.~(\ref{I3}) and (\ref{I4}), 
we see the difference of the gravitational field equations 
in $F(R)$ gravity from those in general relativity. 

\subsection{Explicit form of $F(R)$}

We derive an explicit form of $F(R)$ realizing a domain wall solution. 
For simplicity, we consider the case in which there is no matter. 

We assume that $u$ is given by a function of $y$, $u=u(y)$. 
For example, we take Eq.~(\ref{I8}), for which 
a domain wall can be realized at $y \sim 0$ as shown in Sec.~III. 
By using Eq.~(\ref{eq:4.6}), we can solve $y$ as a function of $R$, $y=y(R)$, 
and therefore we have $u=u(y(R))$. 
Substituting this expression into Eqs.~(\ref{eq:4.4}) and (\ref{eq:4.5}) 
and eliminating $y$, Eqs.~(\ref{eq:4.4}) and (\ref{eq:4.5}) can be rewritten as differential equations for $F(R)$ as a function of $R$. 
Since Eqs.~(\ref{eq:4.4}) and (\ref{eq:4.5}) are not independent 
with each other, we examine Eq.~(\ref{eq:4.4}). 
As a consequence, 
Eq.~(\ref{eq:4.4}) can be expressed as 
\begin{equation} 
{\Xi}_1 (R) \frac{d^2 F(R)}{d R^2} + {\Xi}_2 (R) \frac{d F(R)}{d R} 
- F(R) = 0\,, 
\label{eq:4.11}
%
%
\end{equation}
where 
\begin{eqnarray}
{\Xi}_1 (R) \Eqn{\equiv} \left( d-1 \right) u' \frac{d R}{d y} 
= \left( d-1 \right) \left( \frac{d R}{d y} \right)^2 
\frac{d u(y(R))}{d R}\,, 
\label{eq:4.12} \\ 
%
%
{\Xi}_2 (R) \Eqn{\equiv} 
\left( - d \right) \left[ u'' + \frac{1}{2}\left(u'\right)^2 \right] 
= \left( - d \right) \Biggl[ \frac{d^2 R}{d y^2} \frac{d u(y(R))}{d R} 
+ \left( \frac{d R}{d y} \right)^2 \frac{d^2 u(y(R))}{d R^2} 
\nonumber \\
&&
\hspace{35mm}
{}+ \frac{1}{2} \left( \frac{d R}{d y} \right)^2 
\left( \frac{d u(y(R))}{d R} \right)^2 \Biggr]\,. 
\label{eq:4.13} 
%
%
\end{eqnarray}

We solve Eq.~(\ref{eq:4.6}) in terms of $y$. 
We define $Y \equiv y^2/y_0^2$. 
For $Y = y^2/y_0^2 \ll 1$, we expand exponential terms and take the first 
leading terms in terms of $Y$. As a result, we obtain 
%
\begin{eqnarray} 
Y \Eqn{\equiv} \frac{y^2}{y_0^2} \approx \frac{R - \gamma_1}{\gamma_2}\,, 
\label{eq:4.14} \\
\gamma_1 \Eqn{\equiv} 2d \frac{u_0}{y_0^2} + \frac{d\left(d-1\right)}{l^2}\,,
\label{eq:4.15} \\
\gamma_2 \Eqn{\equiv} -d \frac{u_0}{y_0^2} \left[ 6+\left(1+d\right) u_0 
\right] + \frac{d\left(d-1\right)}{l^2} u_0 \,, 
\label{eq:4.16}
\end{eqnarray}
%
where $\gamma_1$ and $\gamma_2$ are constants. 

By using Eq.~(\ref{eq:4.6}) and the similar procedure, we find 
\begin{eqnarray}
\frac{d R}{d y} \Eqn{\approx} \zeta_1 + \zeta_2 \frac{y^2}{y_0^2}\,, 
\label{eq:4.17} \\
\zeta_1 \Eqn{\equiv} d \frac{u_0}{y_0^3} 
\left( 1 + \frac{1+d}{2} u_0 \right) 
+ \frac{d\left(d-1\right)}{l^2} \frac{u_0}{y_0}\,,
\label{eq:4.18} \\
\zeta_2 \Eqn{\equiv} -d \frac{u_0}{y_0^3} \left[ 1+\left(1+d\right) u_0 
\right] - \frac{d\left(d-1\right)}{l^2} 
\frac{u_0 \left( u_0 + 1 \right)}{y_0}\,, 
\label{eq:4.19} \\ 
\frac{d^2 R}{d y^2} \Eqn{\approx} \eta_1 + \eta_2 \frac{y^2}{y_0^2}\,, 
\label{eq:4.20} \\
\eta_1 \Eqn{\equiv} -d \frac{u_0}{y_0^4} 
\left[ 1 + \left(1+d\right) u_0 \right] 
- \frac{d\left(d-1\right)}{l^2} 
\frac{u_0 \left( 1 - u_0 \right) \e^{-u_0}}{y_0^2}\,,
\label{eq:4.21} \\
\eta_2 \Eqn{\equiv} d \frac{u_0}{y_0^4} 
\left[ 1 + 2\left(1+d\right) u_0 \right] 
+\frac{d\left(d-1\right)}{l^2} 
\frac{u_0 \left( u_0^2 -3u_0 +1 \right)}{y_0^2}\,. 
\label{eq:4.22}
\end{eqnarray}
Here, $\zeta_1$, $\zeta_2$, $\eta_1$ and $\eta_2$ are constants. 

Substituting Eqs.~(\ref{eq:4.17}) and (\ref{eq:4.20}) into 
Eqs.~(\ref{eq:4.12}) and (\ref{eq:4.13}), 
expanding exponential terms, and taking the first 
leading terms in terms of $Y$, 
we acquire 
\begin{equation}  
\Xi_i (R) = {\Xi}^{(0)}_i + {\Xi}^{(1)}_i Y 
= {\Xi}^{(0)}_i -\frac{{\gamma}_1}{{\gamma}_2} {\Xi}^{(1)}_i 
+ {\Xi}^{(1)}_i R\,, 
\label{eq:4.23} 
\end{equation}
with 
%
%
\begin{eqnarray}
{\Xi}^{(0)}_1 \Eqn{\equiv} \left( d-1 \right) 
\left( -  \frac{u_0}{\gamma_2} \zeta_1^2 \right)\,,
\label{eq:4.24} \\ 
{\Xi}^{(1)}_1 \Eqn{\equiv} \left( d-1 \right) \frac{u_0}{\gamma_2} \zeta_1 
\left( \zeta_1 - 2 \zeta_2 \right)\,,
\label{eq:4.25} \\ 
{\Xi}^{(0)}_2 \Eqn{\equiv} \left( - d \right) 
\left[ -\frac{u_0}{\gamma_2} \eta_1 
+\frac{u_0}{\gamma_2^2} \zeta_1^2 \left( 1+\frac{u_0}{2} \right) \right]\,,
\label{eq:4.26} \\ 
{\Xi}^{(1)}_2 \Eqn{\equiv} \left( - d \right) 
\left\{
\frac{u_0}{\gamma_2} \left( \eta_1 - \eta_2 \right) 
- \zeta_1 \frac{u_0}{\gamma_2^2} 
\left[ \zeta_1 \left( 1+u_0 \right) - 2 \zeta_2 
\left( 1+\frac{u_0}{2} \right) \right] 
\right\}\,.
\label{eq:4.27} 
%
%
\end{eqnarray}
In deriving the second equality in Eq.~(\ref{eq:4.23}), we have used 
Eq.~(\ref{eq:4.14}). 
Here, $i, j = 1, 2$, and the superscriptions $(0)$ and $(1)$ denotes 
the terms proportional to the zeroth power of $Y$ ($Y^0 = 1$) and the first 
one of $Y$ ($Y^1 = Y$), respectively. 

For $Y = y^2/y_0^2 \ll 1$, when ${\Xi}^{(1)}_i/{\Xi}^{(0)}_i \lesssim 
\mathcal{O} (1)$, 
we can consider that $\Xi_i \approx {\Xi}^{(0)}_i ( = \mathrm{constant})$. 
In such a case, Eq.~(\ref{eq:4.11}) can be regarded as 
\begin{equation} 
\frac{d^2 F(R)}{d R^2} + \mathcal{C} \frac{d F(R)}{d R} + 
\mathcal{D} F(R) = 0\,, 
\label{eq:4.28} 
\end{equation}
where 
\begin{eqnarray}
\mathcal{C} \Eqn{\equiv} \frac{{\Xi}^{(0)}_2}{{\Xi}^{(0)}_1}\,, 
\label{eq:4.29} \\ 
\mathcal{D} \Eqn{\equiv} -\frac{1}{{\Xi}^{(0)}_1}\,.
\label{eq:4.30} 
%
%
\end{eqnarray}
A general solution of Eq.~(\ref{eq:4.28}) is given by 
\begin{equation} 
F(R) = F_{+} \e^{\lambda_{+} R} + F_{-} \e^{\lambda_{-} R}\,, 
\label{eq:4.31} 
\end{equation}
with 
\begin{equation} 
\lambda_{\pm} \equiv \frac{1}{2} 
\left( -\mathcal{C} \pm \sqrt{{\mathcal{C}}^2-4\mathcal{D}} \right)\,. 
\label{eq:4.32} 
\end{equation}
Here, $F_{\pm}$ are arbitrary constants, and 
the subscriptions $\pm$ of $\lambda_{\pm}$ correspond to the sign 
$\left(\pm\right)$ on the right-hand side (r.h.s.) of Eq.~(\ref{eq:4.32}). 

It follows from the considerations in Sec.~III that 
if we take Eq.~(\ref{I8}), 
the distribution of the energy density is localized at $y \sim 0$ and hence 
a domain wall can be made. Thus, it is interpreted that in 
an exponential model of $F(R)$ gravity given by Eq.~(\ref{eq:4.31}), 
a domain wall can be realized at $y \sim 0$. 
In Ref.~\cite{Exponential-Gravity}, 
such an exponential model of $F(R)$ gravity has been 
studied. 

We should also note that the metric ansatz Eq.~(\ref{I8}) will lead to the 
same ghost and negative energy problem which make the model physically not 
viable as in Sec.~III. 
This is because in this subsection, we examine an explicit form of $F(R)$ 
realizing a domain wall solution constructed in Sec.~III. 

\subsection{Form of the potential in a corresponding scalar field theory}

We examine the form of the potential $V(\phi)$ in Eq.~(\ref{eq:2.5}) 
in a corresponding scalar field theory in the Einstein frame 
to which an exponential model of $F(R)$ gravity in Eq.~(\ref{eq:4.31}) 
is transformed through a conformal transformation in Eq.~(\ref{eq:2.2}). 
As an exponential model, for example, by choosing $F_{+} \neq 0$ and 
$F_{-} = 0$, we take $F(R) = F_{+} \e^{\lambda_{+} R}$. 
In this case, from Eq.~(\ref{eq:2.5}) we have the following relation between 
$R$ and $\phi$: 
\begin{equation}  
R =  \frac{1}{\lambda_{+}} \left[ 
\ln \left( \frac{1}{F_{+} \lambda_{+}} \right) 
+ \sqrt{\frac{2}{3}} \kappa \phi \right]\,. 
\label{eq:4.33} 
\end{equation} 
By using Eqs.~(\ref{eq:2.5}), (\ref{eq:2.7}) and (\ref{eq:4.33}), we find 
\begin{equation} 
V(\phi) = \frac{1}{2 \kappa^2 \lambda_{+}} 
\e^{-\sqrt{2/3} \kappa \phi} 
\left[ \sqrt{\frac{2}{3}} \kappa \phi + 
\ln \left( \frac{1}{F_{+} \lambda_{+}} \right) - 1
\right]\,.  
\label{eq:4.34} 
\end{equation} 
By defining 
$\bar{\phi} \equiv \sqrt{2/3} \kappa \phi$, 
$\bar{\phi}_0 \equiv 
\ln \left[ 1/ \left( F_{+} \lambda_{+} \right) \right] - 1$, 
and 
$V_0 \equiv 1/\left( 2 \kappa^2 \lambda_{+} \right)$, 
$V(\phi)$ in 
Eq.~(\ref{eq:4.34}) is expressed as 
$V(\bar{\phi}) = 
V_0 \e^{-\bar{\phi}} \left( \bar{\phi} + \bar{\phi}_0 \right)$. 
We note that $\bar{\phi}$ is a dimensionless quantity. 

\begin{center}
\begin{figure}[tbp]
   \includegraphics{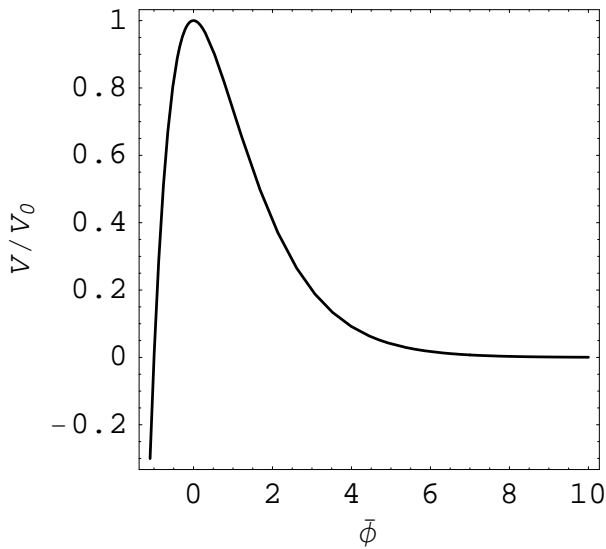}
\caption{$V/V_0$ as a function of $\bar{\phi}$ for 
$\bar{\phi}_0 = 1$ ($F_{+} \lambda_{+} = 1/\e^2$). 
}
\label{Fig-4}
\end{figure}
\end{center}

We show $V/V_0$ as a function of $\bar{\phi}$ in Fig.~\ref{Fig-4} for 
$\bar{\phi}_0 = 1$, i.e., $F_{+} \lambda_{+} = 1/\e^2$. 
From Fig.~\ref{Fig-4}, we see that the potential energy is localized 
at $\bar{\phi} \equiv \sqrt{2/3} \kappa \phi \sim 0$. 
However, it should again be cautioned that 
in the Einstein frame with the potential $V(\phi)$ in 
Eq.~(\ref{eq:4.34}), a domain wall is not formed. 
In other words, it is considered that 
the form of the potential $V(\phi)$ in 
Eq.~(\ref{eq:4.34}) drawn in Fig.~\ref{Fig-4} 
is just a corresponding form to realize an $F(R)$ gravity model of 
$F(R) = F_{+} \e^{\lambda_{+} R}$ with realizing 
a static domain wall solution in the Jordan frame. 
The analyses and considerations in this subsection 
correspond to those in Sec.~II B, and 
the direction of the conformal transformation 
(i.e., from the Jordan frame to the Einstein frame)
is the opposite to that (i.e., from the Einstein frame to the Jordan frame) 
in Sec.~II B. 

\vspace{-16mm}

\section{Effective (gravitational) domain wall} 

In this section, we demonstrate that 
there could exist an effective (gravitational) domain wall 
in the framework of $F(R)$ gravity 
by using the reconstruction method of $F(R)$ 
gravity in Ref.~\cite{Reconstruction-F(R)-N-O}.

\subsection{Reconstruction method}

We now consider $F(R)$ model whose action is given by
\be
\label{JGRG7}
S_{F(R)}= \int d^4 x \sqrt{-g} \left( \frac{F(R)}{2\kappa^2} +
\mathcal{L}_\mathrm{matter} \right)\, .
\ee
Here $F(R)$ is an appropriate function of the scalar curvature $R$. 
The action (\ref{JGRG7}) is equivalently rewritten as
\be
\label{PQR1}
S=\int d^4 x \sqrt{-g} 
\left[ 
\frac{1}{2\kappa^2} \left( P(\psi) R 
+  Q(\psi) \right) + \mathcal{L}_\mathrm{matter}
\right]\, .
\ee
Here, $P$ and $Q$ are proper functions of the auxiliary scalar $\psi$.
By the variation over $\psi$, it follows that 
$0=P'(\psi)R + Q'(\psi)$,
which may be solved with respect to $\psi$ as $\psi=\psi(R)$. 
Here, the prime denotes the derivative with respect to the argument of 
the function as $P'(\psi) = d P(\psi)/d \psi$. 
By substituting the obtained expression of $\psi(R)$ into (\ref{PQR1}),
one arrives again at the $F(R)$-gravity:
\be
\label{PQR4}
S=\int d^4 x \sqrt{-g} 
\left(
\frac{F(R)}{2\kappa^2} +
\mathcal{L}_\mathrm{matter}
\right)\, , \quad
F(R)\equiv P(\psi(R)) R + Q(\psi(R))\, .
\ee
For the action (\ref{PQR1}), the variation of the metric gives
\be
\label{PQ1}
0 = \frac{1}{2}g_{\mu\nu} \left( P(\psi) R + Q(\psi) \right) 
 - P(\psi) R_{\mu\nu} + \nabla_\mu \nabla_\nu P(\psi) - g_{\mu\nu} \Box P(\psi)\, .
\ee
Here we have neglected the contribution from the matter. 

We now assume the 
$D=d+1$ dimensional warped metric in Eq.~(\ref{I2}) 
and we also assume the scalar field only depends on $y$. 
Then $(y,y)$ and $(i,j)$ components of (\ref{PQ1}) give the follwoing equations:
\bea
\label{FRDW1}
0 
\Eqn{=} 
\frac{1}{2} 
\left\{ 
P(\psi) 
\left[ 
- d u'' - \frac{d(d+1)}{4}\left( u' \right)^2 
+ \frac{d(d-1)\e^{-u}}{l^2} 
\right] + Q(\psi) 
\right\}\nn
&& - P(\psi) 
\left[ 
- \frac{d}{2} u'' - \frac{d}{4} \left( u' \right)^2 
\right] - \frac{d-1}{2}u'\psi' P'(\psi) \, , \\
\label{FRDW2}
0 
\Eqn{=} 
\frac{1}{2} \e^u \left\{ P(\psi) 
\left[ 
- d u'' - \frac{d(d+1)}{4}\left( u' \right)^2 
+ \frac{d(d-1)\e^{-u}}{l^2} 
\right] + Q(\psi) \right\} 
\nonumber \\ 
&&
{}- P(\psi) \left\{ \frac{d-1}{l^2} 
+ \e^u 
\left[ 
- \frac{1}{2} u'' - \frac{d}{4} \left( u' \right)^2 
\right] \right\} \nn
&& + \frac{1}{2}\e^u u' \psi' P'(\psi) 
 - \e^u 
\left[ \psi'' P'(\psi) + \left(\psi'\right)^2 P'' (\psi) 
+ \frac{d-1}{2}u' \psi' P'(\psi) 
\right] \, ,
\eea
where $u' = d u(y)/d y$ and $u'' = d^2 u(y)/d y^2$. 
By choosing $\psi=y$, 
in case $1/l^2 =0$, by rewriting Eqs.~(\ref{FRDW1}) and (\ref{FRDW2}), we obtain, 
\bea
\label{FRDW3}
0 
\Eqn{=} 
P''(\psi) - \frac{u'(\psi)}{2} P'(\psi) + \frac{(d-1)u''(\psi)}{2} P(\psi)\, , \\ 
\label{FRDW4}
Q(\psi) 
\Eqn{=} 
\frac{d(d-1) \left(u'(\psi)\right)^2}{4} P(\psi) + (d-1) u'(\psi) P'(\psi) \, .
\eea
Equation (\ref{FRDW3}) can be further rewritten as 
\begin{eqnarray} 
\label{FRDW5}
u'(\psi) 
\Eqn{=} - \frac{2}{d-1} P(\psi)^{\frac{1}{d-1}} \int d\psi P(\psi)^{- \frac{d}{d-1}} P''(\psi) 
\nonumber \\ 
\Eqn{=} - \frac{2}{d-1} 
\left[ 
\frac{P'(\psi)}{P(\psi)} + \frac{d}{d-1} 
P(\psi)^{\frac{1}{d-1}} \int d\psi P(\psi)^{- \frac{2d-1}{d-1}} \left(P'(\psi)\right)^2
\right]\, .
\end{eqnarray}
In the second equality in (\ref{FRDW5}), we have used partial integration. 
Furthermore by writing 
\be
\label{FRDW6}
P(\psi) = U(\psi)^{-2(d-1)}\, ,
\ee
we find
\be
\label{FRDW7}
u'(\psi) = \frac{4U'(\psi)}{U(\psi)} - \frac{8d}{U(\psi)^2} \int d\psi U'(\psi)^2 \, .
\ee

As an example, we consider a model
\be
\label{FRDW8}
U(\psi) = U_0 \left( \psi^2 + \psi_0^2 \right)^\chi\, .
\ee
Here $U_0$, $\psi_0$, and $\chi$ are constants. 
Then we find
\be
\label{FRDW9}
u'(\psi) = \frac{2\chi \psi}{\psi^2 + \psi_0^2} 
 - \frac{32 d \chi^2 \psi^{4\chi -1} }{\left( \psi^2 + \psi_0^2 \right)^{2\chi}}\sum_{k=0}^\infty 
\frac{\Gamma\left(2\chi - 1 \right)}{\left(4\chi -1 - 2k\right) \Gamma\left( 2\chi - 1 - k \right) k!}
\left(\frac{\psi_0^2}{\psi^2}\right)^k\, .
\ee
When $\psi=y$ is large, $u'(\psi)$ behaves as 
\be
\label{FRDW10}
u'(\psi) = \left(2\chi - \frac{32 d \chi^2}{4\chi -1} \right) \frac{1}{\psi} 
+ 
\left[ 
-2\chi + \frac{64 d \chi^3}{4\chi - 1} 
 - \frac{64 d \chi^2 \left(\chi - 1\right)}{4\chi - 3} 
\right] \frac{\psi_0}{\psi^2}
+ \mathcal{O} \left( \left(\frac{\psi_0^2}{\psi^2}\right)^2\right) \, .
\ee
Therefore if we choose
\be
\label{FRDW12}
\chi= - \frac{1}{4\left(4d -1\right)}\, ,
\ee
we find
\be
\label{FRDW13}
u'(\psi)  =  \frac{1}{4 (6 d -1)}\frac{\psi_0}{\psi^2}
+ \mathcal{O} \left( \left(\frac{\psi_0^2}{\psi^2}\right)^2\right) \, .
%
%
\ee
Then by imposing the boundary condition that the universe becomes flat ($u\to 0$) when $|y|=|\psi| \to \infty$, 
we find
\be
\label{FRDW13}
u(\psi)  = - \frac{1}{4 (6 d -1)}\frac{\psi_0}{\psi}
+ \mathcal{O} \left( \left(\frac{\psi_0}{\psi}\right)^3\right) \, .
%
%
\ee
Since $u(\psi)$ behaves non-trivially when $\psi = y \sim 0$, we may regard that there could be an effective 
(gravitational) domain wall at $y=0$.  

\begin{center}
\begin{figure}[tbp]
   \includegraphics{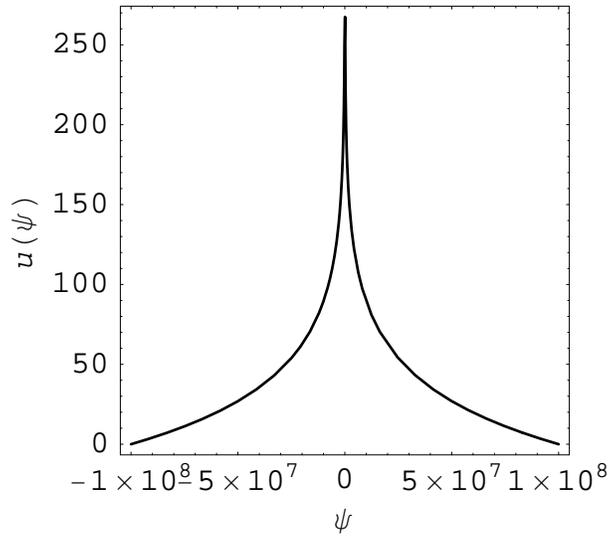}
\caption{$u(\psi)$ as a function of $\psi$ for 
$d=3$, $\chi = 2$, and $\psi_0 = 1$. 
}
\label{Fig-5}
\end{figure}
\end{center}

For a model in Eq.~(\ref{FRDW8}), by using Eq.~(\ref{FRDW7}), 
$u(\psi)$ can be described as 
\begin{equation} 
u(\psi) = 8 \chi \int_{-\infty}^{\psi} d \psi \frac{\psi}{\psi^2 + \psi_0^2} 
-32 d \chi^2 \int_{-\infty}^{\psi} d \psi \frac{1}{\left( \psi^2 + \psi_0^2 
\right)^{2 \chi}} \int_{0}^{\psi} d \tilde{\psi} 
\left( \tilde{\psi}^2 + \psi_0^2 \right)^{2 \left( \chi - 1 \right)} 
\tilde{\psi}^2 \,.
\label{eq:5A-U-psi}
\end{equation}
In Fig.~\ref{Fig-5}, 
we illustrate $u(\psi)$ in Eq.~(\ref{eq:5A-U-psi}) as a function of $\psi$ for 
$d=3$, $\chi = 2$, and $\psi_0 = 1$. 
{}From Fig.~\ref{Fig-5}, we see that 
$u(\psi)$ has a local maximum around $\psi = y \sim 0$, 
and hence it is considered that there could exist an effective 
(gravitational) domain wall at $y=0$. 
In the numerical analysis of Eq.~(\ref{eq:5A-U-psi}) in Fig.~\ref{Fig-5}, 
we have substituted the minimum of $\psi$ in the 
integral range $\psi_{\mathrm{min}} = - 10^{8}$ for $-\infty$. 
We have also checked that 
the qualitative behavior of $u(\psi)$ as a function of $\psi$ 
does not depend on these values of parameters sensitively. 

\vspace{-15mm}

\begin{center}
\begin{figure}[tbp]
   \includegraphics{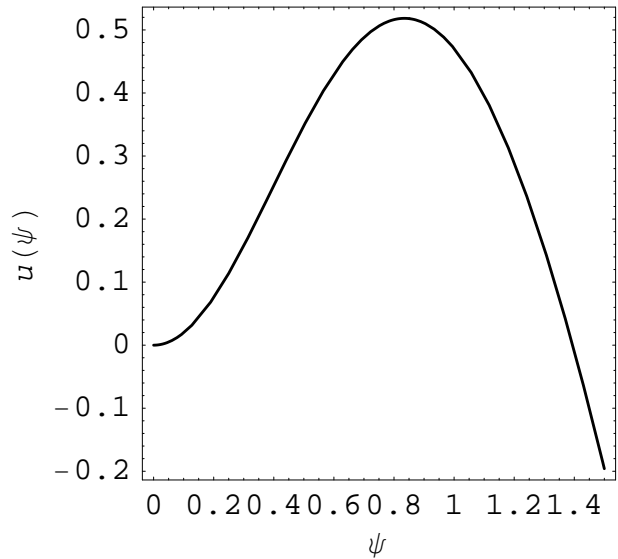}
\caption{$u(\psi)$ in Eq.~(\ref{eq:Addition-5-1}) as a function of $\psi$ for 
$d=3$, $\psi_0 = 1$ and $c_0 = 0$. 
}
\label{Fig-6}
\end{figure}
\end{center}

Furthermore, 
for $\chi = 1/2$ in a model in Eq.~(\ref{FRDW8}), it is possible to 
acquire an analytic solution as follows. 
\begin{equation} 
u(\psi) = 2\left( 1-2d \right) \ln \left( \psi^2 + \psi_0^2 \right) +4d 
\left( \arctan \left( \frac{\psi}{\sqrt{\psi_0^2}} \right) \right)^2 + c_0\,, 
\label{eq:Addition-5-1}
\end{equation}
where $c_0$ is an integration constant. 
We illustrate the behavior of $u(\psi)$ in Eq.~(\ref{eq:Addition-5-1}) 
for $d=3$ (i.e., $D=4$ dimension), $\psi_0 = 1$ and $c_0 = 0$ in 
Fig.~\ref{Fig-6}. 
{}From Fig.~~\ref{Fig-6}, we see that 
$u(\psi)$ has a local maximum around $\psi \sim 0.8$, 
although $u(\psi)$ does not asymptotically approaches $0$ in the limit 
of $|\psi| \to \infty$. 
Thus, it might be interpreted that 
in the range of $|\psi| \lesssim 1.4$, i.e., a small amplitude of $\psi$, 
the distribution of the energy density is localized and hence 
such a configuration could be regarded as 
an effective (gravitational) domain wall. 

\vspace{-16mm}

\subsection{Reconstruction of an explicit form of $F(R)$}

For $u(\psi)$ in Eq.~(\ref{eq:Addition-5-1}), we explicitly derive 
a form of $F(R)$. 
We note that as a possible analytic solution, 
we here consider $u(\psi)$ in Eq.~(\ref{eq:Addition-5-1}), 
even though only in a small amplitude of $\psi$, 
the distribution of the energy density might correspond to 
an effective (gravitational) domain wall. 

By using $0=P'(\psi)R + Q'(\psi)$ and Eq.~(\ref{FRDW4}), we find 
\begin{equation}  
R = -\frac{Q'(\psi)}{P'(\psi)} = -\left( d-1 \right) 
\left( 
\frac{d}{2} \frac{u'(\psi) u''(\psi)}{P'(\psi)} + u''(\psi) + u'(\psi) 
\frac{P''(\psi)}{P'(\psi)} \right)\,. 
\label{eq:Addition-5-2}
\end{equation}

{}From Eq.~(\ref{eq:Addition-5-2}), we derive an analytic relation 
$\psi = \psi(R)$. By substituting this relation into the second 
equation in Eq.~(\ref{PQR4}), we can obtain an explicit form of $F(R)$.  
We define $\bar{Y} \equiv \psi^2/ \psi_0^2$. 
For $\bar{Y} = \psi^2/ \psi_0^2 \ll 1$, we expand each 
quantities in terms of $\bar{Y}$ 
and take leading terms in terms of $\bar{Y}$. 
For $u(\psi)$ in Eq.~(\ref{eq:Addition-5-1}), 
from Eqs.~(\ref{FRDW6}) and (\ref{FRDW8}) with $\chi = 1/2$, 
$P(\psi)$ is described as 
$P(\psi) = \left( U_0 \psi_0 \right)^{-2\left( d-1 \right)} 
\left( 1+\bar{Y} \right)^{-2\left( d-1 \right)}$.
{}From Eq.~(\ref{eq:Addition-5-2}), 
we obtain 
%
\begin{eqnarray} 
R \Eqn{=} {\mathcal{R}}_0 + {\mathcal{R}}_1 \bar{Y}\,, 
\label{eq:Addition-5-3} \\ 
{\mathcal{R}}_0 \Eqn{\equiv} 
\frac{4\left( d-1 \right)}{\psi_0^2} 
\left[ \frac{d}{\left( d-1 \right)}\frac{1}{
\left( U_0 \psi_0 \right)^{-2\left( d-1 \right)}} -2 \right]\,,
\label{eq:Addition-5-4} \\ 
{\mathcal{R}}_1 \Eqn{\equiv} 
-\frac{4\left( d-1 \right)}{\psi_0^2} 
\left[ \frac{d}{\left( d-1 \right)}\frac{1}{ 
\left( U_0 \psi_0 \right)^{-2\left( d-1 \right)}} 
\left( 4+\frac{5d}{3} \right) 
- \left( 4+\frac{14d}{3} \right) 
\right]\,. 
\label{eq:Addition-5-5}
\end{eqnarray}
%
where ${\mathcal{R}}_0$ and ${\mathcal{R}}_1$ are constants. 
Moreover, we have 
%
\begin{eqnarray} 
Q \Eqn{=} {\mathcal{Q}}_1 \bar{Y} + {\mathcal{Q}}_2 \bar{Y}^2\,, 
\label{eq:Addition-5-6} \\ 
{\mathcal{Q}}_1 \Eqn{\equiv} 
\frac{4\left( d-1 \right)}{\psi_0^2} 
\left[ d -2\left( d-1 \right) 
\left( U_0 \psi_0 \right)^{-2\left( d-1 \right)} \right]\,,
\label{eq:Addition-5-7} \\ 
{\mathcal{Q}}_2 
\Eqn{\equiv} 
\frac{8\left( d-1 \right)}{\psi_0^2} 
\left[ -d\left( 1+\frac{2d}{3} \right) 
+ \left( U_0 \psi_0 \right)^{-2\left( d-1 \right)} 
\left( d-1 \right) \left( 1+\frac{5d}{3} \right) 
\right]\,, 
\label{eq:Addition-5-8}
\end{eqnarray}
%
where ${\mathcal{Q}}_1$ and ${\mathcal{Q}}_2$ are constants. 
By using Eq.~(\ref{eq:Addition-5-3}), we express $\bar{Y}$ as 
%
\begin{eqnarray} 
\bar{Y} \Eqn{=} \bar{Y}_0 + \bar{Y}_1 R\,, 
\label{eq:Addition-5-9} \\ 
\bar{Y}_0 
\Eqn{\equiv} 
-\frac{{\mathcal{R}}_0}{{\mathcal{R}}_1}\,, 
\label{eq:Addition-5-10} \\ 
\bar{Y}_1 
\Eqn{\equiv} 
\frac{1}{{\mathcal{R}}_1}\,. 
\label{eq:Addition-5-11}
\end{eqnarray}
%
We expand $P(\psi)$ as 
$P(\psi) \approx \left( U_0 \psi_0 \right)^{-2\left( d-1 \right)} 
\left\{ 1-\left( d-1 \right)\bar{Y} + 
\left[d\left( d-1 \right)/2\right]\bar{Y}^2  \right\}$. 
We substitute this relation and Eq.~(\ref{eq:Addition-5-6}) with 
Eq.~(\ref{eq:Addition-5-9}) into the second 
equation in Eq.~(\ref{PQR4}) and 
take terms which is of order of $R^2$. 
As a consequence, we acquire
%
\begin{eqnarray} 
F(R) \Eqn{=} {\mathcal{F}}_0 + {\mathcal{F}}_1 R + {\mathcal{F}}_2 R^2 \,, 
\label{eq:Addition-5-12} \\ 
{\mathcal{F}}_0 \Eqn{\equiv} 
{\mathcal{Q}}_1 \bar{Y}_0 + {\mathcal{Q}}_2 \bar{Y}_0^2 \,,
\label{eq:Addition-5-13} \\ 
{\mathcal{F}}_1 \Eqn{\equiv} 
\left( U_0 \psi_0 \right)^{-2\left( d-1 \right)} 
\left[ 
1-\left( d-1 \right) \bar{Y}_0 + \frac{d\left( d-1 \right)}{2} \bar{Y}_0^2
\right] 
+ {\mathcal{Q}}_1 \bar{Y}_1 + 2 {\mathcal{Q}}_2 \bar{Y}_0 \bar{Y}_1 \,,
\label{eq:Addition-5-14} \\ 
{\mathcal{F}}_2 
\Eqn{\equiv} 
\left( U_0 \psi_0 \right)^{-2\left( d-1 \right)} 
\left( d-1 \right) \bar{Y}_1 \left( -1 + d\bar{Y}_0 \right) 
+ {\mathcal{Q}}_2 \bar{Y}_1^2 \,,
\label{eq:Addition-5-15} 
\end{eqnarray}
%
where ${\mathcal{F}}_0$, ${\mathcal{F}}_1$ and ${\mathcal{F}}_2$ 
are constants. 
Since we have derived an explicit form of $F(R)$ 
in Eq.~(\ref{eq:Addition-5-12}) for $\bar{Y} = \psi^2/ \psi_0^2 \ll 1$, 
from Eq.~(\ref{eq:Addition-5-3}) it can be considered that 
this $F(R)$ form in Eq.~(\ref{eq:Addition-5-12}) corresponds to the one 
for $R \sim \mathcal{O} (1)$ 
when ${\mathcal{R}}_0 \sim \mathcal{O} (1)$. 
If we set ${\mathcal{F}}_0 = 0$ and ${\mathcal{F}}_1 = 1$, 
from Eq.~(\ref{eq:Addition-5-15}) we find 
$F(R) = R + {\mathcal{F}}_2 R^2$. 
In the limit of the small curvature regime, 
$F(R)$ asymptotically approaches $R$, i.e., general relativity. 
Thus, for $u(\psi)$ in Eq.~(\ref{eq:Addition-5-1}) forming an effective 
(gravitational) domain wall, 
an explicit form of $F(R)$ is described by a power-law model. 

Finally, we clearly explain 
the difference between the investigations for $F(R)$ gravity with 
a static domain wall solution in Sec.~IV and 
the demonstrations for an effective (gravitational) domain wall 
in $F(R)$ gravity in Sec.~V. 
In Sec.~IV, 
we regard the deviation of $F(R)$ gravity from general relativity 
as a geometrical contribution to the energy-momentum tensor, 
which can be described in Eq.~(\ref{eq:4.2}). 
Since we have a static domain wall solution in a scalar field theory 
in general relativity in Sec.~III, 
by comparing Eqs.~(\ref{eq:4.7}) and (\ref{eq:4.8}) with 
Eqs.~(\ref{I3}) and (\ref{I4}), 
we find the difference of the gravitational field equations 
in $F(R)$ gravity from those in general relativity. 
Furthermore, in principle 
we can derive an explicit form of $F(R)$ realizing a domain wall 
solution, as discussed in Sec.~IV B. 
In other words, in Sec.~IV 
we first suppose the existence of a static domain wall solution 
in $F(R)$ gravity, which is equivalent to that obtained 
in a scalar field theory in general relativity in Sec.~III. 
Then, through the comparison of gravitational field equations in 
$F(R)$ gravity with those in a scalar field theory in general relativity, 
we reconstruct an explicit form of $F(R)$. 
On the other hand, in Sec.~V, 
by using the reconstruction method of 
$F(R)$ gravity~\cite{Reconstruction-F(R)-N-O}, 
we directly show that 
the distribution of the energy density could be localized 
and hence such a configuration could be regarded as 
an effective (gravitational) domain wall. 
Here, the reason why we call ``an effective (gravitational) domain wall'', 
i.e., what is the definition of it, 
is that a domain wall solution obtained in Sec.~V 
is realized by a pure gravitational effect. 
This is because in Sec.~V 
we consider the case in which there is no matter, such as a scalar field, 
whereas the realization of a static domain wall solution explored in Sec.~III 
comes from the existence of a scalar field $\varphi$ in the action 
in Eq.~(\ref{I1}). 
As a result, 
the fundamental difference of an effective (gravitational) domain wall 
investigated in Sec.~V from the domain wall solution obtained in Sec.~IV 
is summarized as follows. 
An effective (gravitational) domain wall in Sec.~V 
is realized by a pure gravitational effect. 
On the other hand, 
a static domain wall solution, the existence of which is shown in Sec.~III, 
is made by a scalar field. 
In Sec.~IV, the deviation of $F(R)$ gravity from general relativity 
contributes the energy-momentum tensor geometrically, 
and eventually it plays an equivalent role of matter, such as a scalar field 
in Sec.~III. 

\section{Non-minimal Maxwell-$F(R)$ gravity} 

It is known that a coupling between the scalar curvature
and the electromagnetic field arises in curved space-time 
due to one-loop vacuum-polarization effects in Quantum Electrodynamics
(QED)~\cite{Drummond:1979pp}. 
Therefore, in this section, as a possible way to examine 
a connection between $F(R)$ gravity 
and variation of the fine structure constant, 
we investigate non-minimal Maxwell-$F(R)$ gravity. 
This might lead to a clue to solve an issue of 
variation of the fine structure constant. 
Furthermore, a non-minimal gravitational coupling of the electromagnetic field 
breaks the conformal invariance of the electromagnetic field. 

\subsection{Variation of the fine structure constant}

We study a case in which there exists a non-minimal 
gravitational coupling of the electromagnetic field 
in $F(R)$ gravity~\cite{Bamba:2008ja}. 
Cosmological consequences of such a non-minimal gravitational 
coupling of the Maxwell field~\cite{Bamba:2008ut} and 
a non-minimal gravitational coupling of the Yang-Mills 
field~\cite{Bamba:2008xa} have also been studied. 

We consider the following action~\cite{Bamba:2008ja}: 
\begin{equation} 
S = \int d^4 x \sqrt{-g} 
\frac{F(R)}{2\kappa^2} 
+ 
\int d^4 x \sqrt{-g} 
\left(
-\frac{1}{4} I(R) 
g^{\mu\alpha} g^{\nu\beta} F_{\mu\nu} F_{\alpha\beta} 
\right)\,, 
\label{eq:3.1}
\end{equation}
where 
\begin{equation} 
I(R) = 1 + \mathcal{I}(R)\,. 
\label{eq:3.2}
\end{equation}
Here, $\mathcal{I}(R)$ is an arbitrary function of $R$. 

We investigate a situation in which a domain wall as well as 
the variation of the fine structure constant can be realized 
in non-minimal Maxwell-$F(R)$ gravity. 
As an $F(R)$ gravity model to form a domain wall, 
we take $F(R) = F_{+} \e^{\lambda_{+} R}$ as in Sec.~IV C. 
Moreover, we choose a logarithmic non-minimal gravitational coupling of 
the electromagnetic field as
\begin{equation} 
I(R) = 1 + \ln \left( \frac{R}{R_0} \right)\,, 
\label{eq:5.5}
\end{equation}
where $R_0$ is the current curvature. 
(Here, $\mathcal{I}(R) = \ln \left( R/R_0 \right)$.) 
In Ref.~\cite{Elizalde:1996am}, 
it has been found that such a logarithmic-type non-minimal gravitational 
coupling appears in the effective renormalization-group improved Lagrangian 
for an $SU(2)$ gauge theory in matter sector for a de Sitter background. 
This comes from the running gauge coupling constant with 
the asymptotic freedom in a non-Abelian gauge theory, which 
approaches zero in very high energy regime. 

Furthermore, from the second part of the action in Eq.~(\ref{eq:3.1}) 
describing non-minimal electromagnetic field theory 
we find 
\begin{equation} 
{\alpha}_{\mathrm{EM}} (R) = 
\frac{{\alpha}_{\mathrm{EM}}^{(0)}}{I(R)}\,, 
\label{eq:5.6}
\end{equation}
where ${\alpha}_{\mathrm{EM}}^{(0)}$ is the bare fine structure constant 
and hence ${\alpha}_{\mathrm{EM}}^{(0)} = {\alpha}_{\mathrm{EM}} (R_0)$. 
Since $R$ is large in the early universe and it decreases in time 
as the universe evolves, ${\alpha}_{\mathrm{EM}}$ varies in time. 
For a logarithmic-type non-minimal gravitational 
coupling in Eq.~(\ref{eq:5.5}), 
we see that 
${\alpha}_{\mathrm{EM}}$ increases as the universe evolves and 
approaches the value of the bare fine structure constant 
at the present time. 

It is known that 
there exist strong constraints on variation of the fine structure constant 
from the big-bang nucleosynthesis (BBN) at redshift $z \sim 1 \times \e^{10}$ 
and from the primary CMB signal. 
Furthermore, there are astronomical constraints from quasar 
absorption lines. 
Moreover, the start formation could be affected by a time-varying 
${\alpha}_{\mathrm{EM}}$ as well.

According to the latest results of Keck/HIRES (High Resolution Echelle 
Spectrometer) quasi-stellar object (QSO) absorption spectra 
over the redshift range $0.2 < z_{\mathrm{abs}} < 3.7$ 
in Ref.~\cite{Murphy:2003hw}, 
${\alpha}_{\mathrm{EM}}$ was smaller in the past and 
the following weighted mean ${\alpha}_{\mathrm{EM}}$ 
with raw statistical errors has been presented: 
\begin{equation} 
\frac{{\alpha}_{\mathrm{EM}} - {\alpha}_{\mathrm{EM}}^{(0)}}
{{\alpha}_{\mathrm{EM}}^{(0)}} 
= \left(-0.543 \pm 0.116 \right) \times 10^{-5}\,,
\label{eq:5.7}
\end{equation}
representing $4.7 \sigma$ significance level. 
For a logarithmic-type non-minimal gravitational 
coupling in Eq.~(\ref{eq:5.5}), we see that 
${\alpha}_{\mathrm{EM}}$ was smaller in the past 
and becomes larger in time. 

In addition, in Ref.~\cite{Webb:2010hc}, 
by analyzing the combined dataset from 
the Keck telescope and 
the ESO Very Large Telescope (VLT), 
the following spatial variation of the fine structure constant 
has been given: 
\begin{equation} 
\frac{{\alpha}_{\mathrm{EM}} - {\alpha}_{\mathrm{EM}}^{(0)}}
{{\alpha}_{\mathrm{EM}}^{(0)}} 
= \left(1.10 \pm 0.25 \right) \times 10^{-6} r \cos \Theta \, 
\mathrm{Glyr}^{-1}\,,
\label{eq:5.8}
\end{equation}
with a significance of $4.2 \sigma$. 
Here, $r(z) \equiv c t(z)$ with $c$ being the speed of light 
is the look-back time at redshift $z$ and 
$\Theta$ is the angle on the sky between sightline and best-fit 
dipole position. 
In Ref.~\cite{Webb:2010hc}, 
by using a new dataset from the ESO VLT, 
it has also been mentioned that 
${\alpha}_{\mathrm{EM}}$ appears on average to be larger in the past. 

It should be cautioned that in our model 
it is not possible to estimate the spatial variation of 
${\alpha}_{\mathrm{EM}}$ and only the time-variation of alpha 
could be estimated. 

For a logarithmic non-minimal gravitational coupling of 
the electromagnetic field in Eq.~(\ref{eq:5.5}), 
in order to compare the theoretical results with the observations on 
the time variation of the fine structure constant 
from quasar absorption lines in Eq.~(\ref{eq:5.7}), 
we estimate the time variation of the fine structure constant 
from the epoch of the redshift $z = 0.21$ to the present time. 
In the flat FLRW space-time in Eq.~(\ref{eq:2.18}), from 
$R/R_0 \approx \left( 1+z  \right)^3$~\cite{Kolb and Turner} 
we find $R(z = 0.21)/R_0 \approx 1.77$. 
By using Eqs.~(\ref{eq:5.5}) and (\ref{eq:5.6}), we obtain 
\begin{equation} 
\frac{{\alpha}_{\mathrm{EM}} (R(z = 0.21)) - {\alpha}_{\mathrm{EM}}^{(0)}}
{{\alpha}_{\mathrm{EM}}^{(0)}} 
= -0.364\,.
\label{eq:5.9}
\end{equation}
This implies that the naive model of a logarithmic non-minimal gravitational 
coupling of the electromagnetic field could not satisfy 
the constraints on the time variation of the fine structure constant 
from quasar absorption lines in Eq.~(\ref{eq:5.7}) 
and therefore it would be ruled out. 

We remark that the time variation of the fine structure constant in the 
Jordan frame depends only on a non-minimal 
gravitational coupling of the electromagnetic field, i.e., 
the form of $I(R)$, and it does not on the form of $F(R)$, 
provided that there is no explicit relation between the form of $F(R)$ and 
that of $I(R)$ in the action in Eq.~(\ref{eq:3.1}). 
In the next subsections, therefore, we explore the effect of $F(R)$ gravity 
on variation of the fine structure constant by 
making a conformal transformation to the Einstein frame. 

\subsection{Relation to a coupling between the electromagnetic field 
and a scalar field in the Einstein frame}

We study the effect of $F(R)$ gravity with realizing a domain wall 
on variation of the fine structure constant. 
By using the same procedure presented in Sec.~II A, 
we make a conformal transformation to the Einstein frame 
in Eq.~(\ref{eq:2.2}). 
Consequently, we obtain 
the action in the Einstein frame described as 
\begin{equation} 
S_{\mathrm{E}} = 
\int d^4 x \sqrt{-\tilde{g}} \left( \frac{\tilde{R}}{2\kappa^2} - 
\frac{1}{2} \tilde{g}^{\mu\nu} {\partial}_{\mu} \phi {\partial}_{\nu} \phi 
- V(\phi) \right) +
\int d^4 x \sqrt{-\tilde{g}} 
\left(
-\frac{1}{4} J(\phi) \tilde{g}^{\mu\alpha} \tilde{g}^{\nu\beta}
F_{\mu\nu} F_{\alpha\beta} 
\right)\,, 
\label{eq:3.3}
\end{equation}
where 
%
\begin{eqnarray}
J(\phi) \Eqn{\equiv} 
e^{-2/\sqrt{3} \kappa \phi} \left( I(\tilde{R}) - 
\frac{d I(\tilde{R})}{d \tilde{R}} \tilde{R} \right) 
\nonumber \\ 
&& 
{}+ e^{-1/\sqrt{3} \kappa \phi} 
\frac{d I(\tilde{R})}{d \tilde{R}} 
\left[ \tilde{R} + \sqrt{3} \tilde{\Box} \left( \kappa \phi \right) 
- \frac{1}{2} \tilde{g}^{\mu\nu}
{\partial}_{\mu} \left( \kappa \phi \right) 
{\partial}_{\nu} \left( \kappa \phi \right)
\right]\,. 
\label{eq:3.4}
\end{eqnarray}
Here, the first term on the r.h.s. of Eq.~(\ref{eq:3.3}) 
is the same as that in Eq.~(\ref{eq:2.6}). 
We note that if $\tilde{R}$ can be expressed by $\phi$, 
$J$ can be described as a function of $\phi$. 
By comparing the second term on the r.h.s. of Eq.~(\ref{eq:3.3}) with 
that of Eq.~(\ref{eq:2.10}), we find $J(\phi) = B(\phi)$. 
Thus, by using this relation, it might be possible that 
we obtain the relation between 
a non-minimal gravitational coupling of the electromagnetic field 
in the Jordan frame and 
a coupling of the electromagnetic field to a scalar field 
in the Einstein frame.

\subsection{Case for an exponential model}

First, 
we take an $F(R)$ gravity model with realizing a domain wall 
as $F(R) = F_{+} \e^{\lambda_{+} R}$ derived in Sec.~IV B 
and a logarithmic non-minimal gravitational coupling of 
the electromagnetic field in Eq.~(\ref{eq:5.5}). 
We also assume the $D=4$ ($d=3$) 
dimensional warped metric in Eq.~(\ref{I2}) because such an 
exponential model $F(R) = F_{+} \e^{\lambda_{+} R}$ is derived 
in this metric in Sec.~IV B. 
We consider the case in which $\phi$ only depends on $y$. 
In this case, 
the effect of $F(R)$ gravity with realizing a domain wall 
is involved in $J(\phi)$ in Eq.~(\ref{eq:3.4}) through the 
relation between the scalar curvature $R$ and $\phi$. 
{}From Eq.~(\ref{eq:3.4}), we obtain 
\begin{equation} 
J(\phi) = 
e^{-2/\sqrt{3} \kappa \phi} \ln \left( \frac{R}{R_0} \right) 
+ e^{-1/\sqrt{3} \kappa \phi} 
\left[ 1-3\sqrt{3} \left( \frac{\phi}{y_0} \right) 
\e^{-\left( \phi/y_0 \right)^2} \frac{u_0}{y_0} \frac{\kappa}{R} 
\left( \frac{d \phi}{d y} \right) - \frac{1}{2} \frac{\kappa^2}{R} 
\left( \frac{d \phi}{d y} \right)^2 
\right]\,, 
\label{eq:Addition-6-C-1}
\end{equation}
%
where $R$ can be described as a function of $\phi$ 
by Eq.~(\ref{eq:4.33}). 
We also take the value of the current curvature 
$R_0 =  \left(1/\lambda_{+}\right) \left\{ 
\ln \left[ 1/\left(F_{+} \lambda_{+} \right) \right] 
+ \sqrt{2/3} \kappa \phi_{\mathrm{p}} \right\}$ by using Eq.~(\ref{eq:4.33}). 
Here, $\phi_{\mathrm{p}}$ is the amplitude of $\phi$ at the present time. 
We note that $R_0$ is determined by $\phi_{\mathrm{p}}$ 
and not $\phi_0$. 

\begin{center}
\begin{figure}[tbp]
   \includegraphics{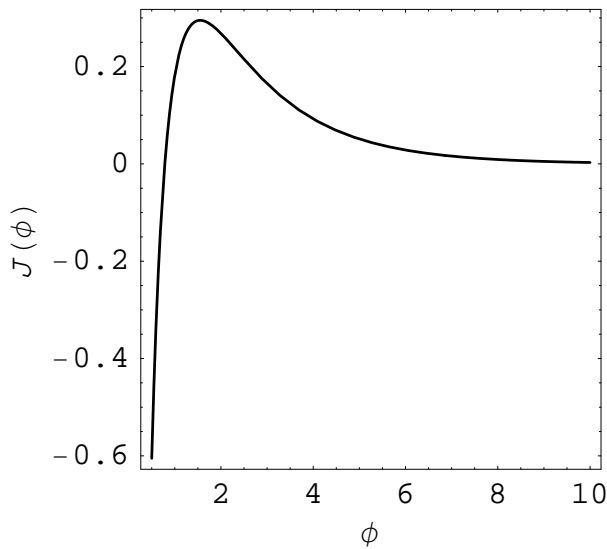}
\caption{$J(\phi)$ as a function of $\phi$ for 
$F_{+} = 1$, $\lambda_{+} = 1$, $u_0 =1$, $y_0 = \phi_0 = 10$, 
and $\phi_{\mathrm{p}} = 1$. 
}
\label{Fig-9}
\end{figure}
\end{center}

We may now choose $\phi=y$ and set $\kappa^2 = 1$ as executed in Sec.~III. 
In Fig.~\ref{Fig-9}, 
we depict $J(\phi)$ 
as a function of $\phi$ for 
$F_{+} = 1$, $\lambda_{+} = 1$, $u_0 =1$, $y_0 = \phi_0 = 10$, 
and $\phi_{\mathrm{p}} = 1$. 
We have confirmed that 
the qualitative behavior of $J(\phi)$ as a function of $\phi$ 
does not depend on these values of parameters sensitively. 

\vspace{-13.8mm}

Moreover, from the second part of the action in Eq.~(\ref{eq:3.3}) 
describing electromagnetic field theory 
we have 
\begin{equation} 
{\alpha}_{\mathrm{EM}} (\phi) = 
\frac{{\alpha}_{\mathrm{EM}}^{(0)}}{J(\phi)}\,, 
\label{eq:Addition-6-C-2}
\end{equation}
where ${\alpha}_{\mathrm{EM}}^{(0)} = {\alpha}_{\mathrm{EM}} 
(\phi_{\mathrm{p}})$. 
We investigate the time variation of the fine structure constant 
from the epoch of the redshift $z = 0.21$ to the present time, 
as executed in Sec.~VI A. 
As an example, 
we choose $F_{+} = 1$, $\lambda_{+} = 1$, $u_0 =1$, $y_0 = \phi_0 = 10$, 
and $\phi_{\mathrm{p}} = 1$. 
Since $R= \sqrt{2/3} \phi$ from Eq.~(\ref{eq:4.33}), we have 
$\phi (z = 0.21) = \left( R(z = 0.21)/R_0 \right) \phi_{\mathrm{p}}$. 
{}From Eq.~(\ref{eq:Addition-6-C-1}), we acquire 
$J(\phi(z = 0.21)) = 0.420$. 
By combining this value with Eq.~(\ref{eq:Addition-6-C-2}), we find 
\begin{equation} 
\frac{{\alpha}_{\mathrm{EM}} (\phi(z = 0.21)) - 
{\alpha}_{\mathrm{EM}}^{(0)}}
{{\alpha}_{\mathrm{EM}}^{(0)}} 
= \frac{1}{J(\phi(z = 0.21))} -1 = 1.38\,.
\label{eq:Addition-6-C-3}
\end{equation}
This value is larger than the constraints on the time variation of the fine 
structure constant from quasar absorption lines in Eq.~(\ref{eq:5.7}) and 
hence the naive model of a logarithmic non-minimal gravitational 
coupling of the electromagnetic field could not be consistent with 
the observations of quasar absorption lines. 
In Fig.~\ref{Fig-9}, we see that 
$J(\phi)$ approaches zero as $\phi$ becomes large. 
It follows from Eq.~(\ref{eq:Addition-6-C-2}) that 
${\alpha}_{\mathrm{EM}}$ decreases as the universe evolves. 
Such a behavior of ${\alpha}_{\mathrm{EM}}$ in the Einstein frame 
is opposite to that in the Jordan frame examined in Sec.~VI A. 

It is interesting to emphasize that in the Einstein frame, the differences of 
$F(R)$ gravity models reflect time-variation of the fine structure 
constant through $J(\phi)$ in Eq.~(\ref{eq:3.4}) due to the 
relation (\ref{eq:2.5}) between $\phi$ and $F_{,R}$. 

\subsection{Case for a power-law model}

Next, 
we take an $F(R)$ gravity model with forming an effective 
(gravitational) domain wall 
as $F(R) = R + {\mathcal{F}}_2 R^2$ derived in Sec.~V B 
and a logarithmic non-minimal gravitational coupling of 
the electromagnetic field in Eq.~(\ref{eq:5.5}). 
We again assume the $D=4$ ($d=3$) 
dimensional warped metric in Eq.~(\ref{I2}) because such a 
power-law model is the one for 
$u(\psi)$ in Eq.~(\ref{eq:Addition-5-1}) derived 
in this metric in Sec.~V A. 
We consider the case in which $\phi$ only depends on $y$. 
In this case, 
the effect of $F(R)$ gravity with forming an effective 
(gravitational) domain wall 
is included in $J(\phi)$ in Eq.~(\ref{eq:Addition-6-C-1}) 
through the following relation between the scalar curvature $R$ and $\phi$: 
\begin{equation}  
R =  \frac{1}{2{\mathcal{F}}_2} \left( \e^{\sqrt{2/3} \kappa \phi} - 1 \right)
\,, 
\label{eq:Addition-6-D-1}
\end{equation}
where we have used Eq.~(\ref{eq:2.5}). 
By using Eq.~(\ref{eq:Addition-6-D-1}), 
we also take the value of the current curvature 
$R_0 = \left( \e^{\sqrt{2/3} \kappa \phi_{\mathrm{p}}} - 1 
\right)/\left( 2{\mathcal{F}}_2 \right)$. 

\begin{center}
\begin{figure}[tbp]
   \includegraphics{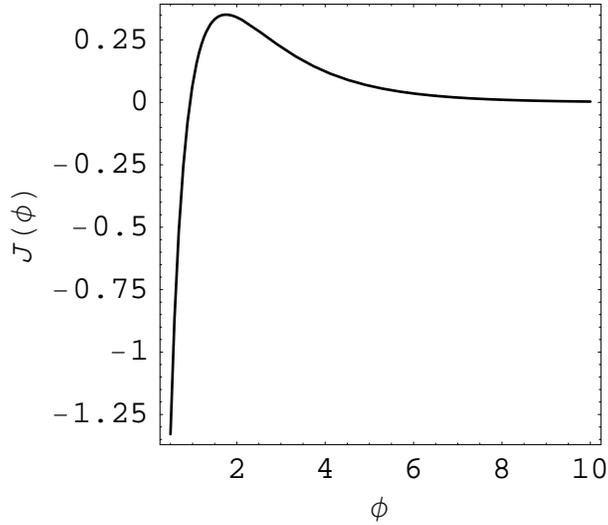}
\caption{$J(\phi)$ as a function of $\phi$ for 
${\mathcal{F}}_2 = 1$, $u_0 =1$, $y_0 = \phi_0 = 10$, and 
$\phi_{\mathrm{p}} = 1$. 
}
\label{Fig-10}
\end{figure}
\end{center}

Here, 
we may choose $\phi=y$ and set $\kappa^2 = 1$ as executed in Sec.~III. 
In Fig.~\ref{Fig-10}, 
we show $J(\phi)$ as a function of $\phi$ for 
${\mathcal{F}}_2 = 1$, $u_0 =1$, $y_0 = \phi_0 = 10$, and 
$\phi_{\mathrm{p}} = 1$. 
We have again verified that 
the qualitative behavior of $J(\phi)$ as a function of $\phi$ 
does not depend on these values of parameters sensitively. 

\vspace{-15.5mm}

We explore the time variation of the fine structure constant 
from the epoch of the redshift $z = 0.21$ to the present time, 
similarly to those in Secs.~VI A and VI C. 
As an example, 
we take ${\mathcal{F}}_2 = 1$, $u_0 =1$, $y_0 = \phi_0 = 10$, and 
$\phi_{\mathrm{p}} = 1$, which is the case illustrated 
in Fig.~\ref{Fig-10}. 
Since $R = \left( \e^{\sqrt{2/3} \phi} - 1 \right)/2$ 
from Eq.~(\ref{eq:Addition-6-D-1}), we acquire 
$\phi(z = 0.21) = 
\sqrt{3/2} \ln \left[ \left( R(z = 0.21)/R_0 \right) 
\left( \e^{\sqrt{2/3}\phi_{\mathrm{p}}} - 1 \right) + 1
\right]$. {}From this relation, for $\phi_{\mathrm{p}} = 1$, 
we obtain $\phi(z = 0.21) = 1.44$. 
By using Eq.~(\ref{eq:Addition-6-C-1}), we find 
$J(\phi(z = 0.21)) = 0.632$. 
By substituting this value into Eq.~(\ref{eq:Addition-6-C-2}), we obtain 
\begin{equation} 
\frac{{\alpha}_{\mathrm{EM}} (\phi(z = 0.21)) - 
{\alpha}_{\mathrm{EM}}^{(0)}}
{{\alpha}_{\mathrm{EM}}^{(0)}} 
= \frac{1}{J(\phi(z = 0.21))} -1 = 0.583\,.
\label{eq:Addition-6-D-2}
\end{equation}
This value is also larger than the constraints on the time variation of 
the fine structure constant from quasar absorption lines in Eq.~(\ref{eq:5.7}) 
and therefore the naive model of a logarithmic non-minimal gravitational 
coupling of the electromagnetic field would be incompatible with 
the observations of quasar absorption lines. 
In Fig.~\ref{Fig-10}, we see that 
$J(\phi)$ approaches zero as $\phi$ increases. 
It follows from Eq.~(\ref{eq:Addition-6-C-2}) that 
${\alpha}_{\mathrm{EM}}$ becomes small as the universe 
evolves, similarly to that for an exponential model in Sec.~VI C. 
Again, such a behavior of ${\alpha}_{\mathrm{EM}}$ in the Einstein frame 
is opposite to that in the Jordan frame examined in Sec.~VI A. 

Finally, we emphasize 
the main reason why we study three types of non-minimal gravitational couplings between the scalar field (or the Ricci scalar) and the electromagnetic field: 
logarithmic, exponential and power law, 
even though these models can be easily ruled out. 
It would be considered that such a non-minimal coupling of the electromagnetic field to gravity could be one of the most theoretically motivated approaches to 
investigate a relation between $F(R)$ gravity and variation of the fine structure constant, and thus that cosmological considerations on this model could 
present an understanding on the origin of variation of the fine structure 
constant. 

\section{Cosmological consequences of the coupling of the electromagnetic 
field to not only a scalar field but also the scalar curvature}

In this section, we consider a scalar field theory with its potential 
forming a domain wall, 
e.g., $V(\phi)$ in Eq.~(\ref{eq:2.9}), 
and its coupling to the electromagnetic field, such as
the action in Eq.~(\ref{eq:2.8}). 
In particular, we extend the coupling of the electromagnetic field 
not only to a scalar field 
but also to the scalar curvature as 
\begin{eqnarray} 
S_{\mathrm{E}} \Eqn{=} 
\int d^4 x \sqrt{-\tilde{g}} \left( \frac{\tilde{R}}{2\kappa^2} - 
\frac{1}{2} \tilde{g}^{\mu\nu} {\partial}_{\mu} \phi {\partial}_{\nu} \phi 
- V(\phi) \right) +
\int d^4 x \sqrt{-\tilde{g}} 
\left(
-\frac{1}{4} \Upsilon(\phi, \tilde{R}) \tilde{g}^{\mu\alpha} 
\tilde{g}^{\nu\beta}F_{\mu\nu} F_{\alpha\beta} \right) 
\nonumber \\ 
&& 
{}+ S_{\mathrm{matter}}\,,
\label{eq:7.1}
\end{eqnarray}
where $\Upsilon(\phi, \tilde{R})$ is an arbitrary function of $\phi$ 
as well as $\tilde{R}$. 
In this case, 
the cosmological evolution of the scalar field $\phi$ as well as 
that of the scalar curvature $\tilde{R}$ can contribute to 
the variation of the fine structure constant. 
Hence, 
a domain wall can be used to account for the 
spatial variation through a scalar field coupled to electromagnetism as in 
Ref.~\cite{Olive:2010vh}, 
whereas the non-minimal gravitational coupling of the electromagnetic field to 
the scalar curvature can explain 
the time variation of the fine structure constant. 
Thus, there exist more choices of the scalar field potential which can make 
a domain wall. 

In addition, it is interesting to remark that 
the conformal invariance of the electromagnetic field can be 
broken by the coupling of the electromagnetic field to 
both a scalar field 
(or a scalar quantity)~\cite{Turner:1987bw, Scalar-EM-Magnetic-fields} and 
the scalar curvature~\cite{Turner:1987bw, Magnetic-fields-Bamba-Sasaki}, and 
therefore large-scale magnetic fields can be generated 
from inflation even in the FLRW spacetime, which is 
conformally flat~\cite{Turner:1987bw, Conformally-flat-FLRWspacetime}\footnote{
In Ref.~\cite{Barrow:2011ic}, 
it has been shown that 
by assuming an open FLRW background, 
large-scale magnetic fields with its enough strength to seed 
the galactic dynamo mechanism can be generated 
within standard electromagnetism and standard general relativity.}
(for a recent review of the generation of primordial magnetic fields, 
see~\cite{Kandus:2010nw}).

Finally, we mention that we can develop the action in Eq.~(\ref{eq:7.1}) 
in the framework of $F(R)$ gravity as follows: 
\begin{eqnarray} 
S \Eqn{=} 
\int d^4 x \sqrt{-g} \left( \frac{F(R)}{2\kappa^2} - 
\frac{1}{2} g^{\mu\nu} {\partial}_{\mu} \phi {\partial}_{\nu} \phi 
- V(\phi) \right) +
\int d^4 x \sqrt{-g} 
\left(-\frac{1}{4} \Upsilon(\phi, R) g^{\mu\alpha} 
g^{\nu\beta}F_{\mu\nu} F_{\alpha\beta} \right) 
\nonumber \\ 
&& 
{}+ S_{\mathrm{matter}}\,. 
\label{eq:7.2}
\end{eqnarray}
In this model action, power-law inflation can occur due to the non-minimal
gravitational coupling of the electromagnetic field as well as 
the deviation of $F(R)$ gravity from general relativity
and the late-time accelerated expansion of the universe can also be 
realized through the modified part of $F(R)$ gravity 
in a unified model action~\cite{Bamba:2008ja, Bamba:2008xa}. 
%
In the scalar-tensor sector of the theory in Eq.~(\ref{eq:7.2}), 
the domain wall may be created due to combined effect of scalar potential and 
modified gravity. Then, combined effect of scalar and curvature in 
the non-minimal electromagnetic sector gives us 
wider possibility for realizing the time-variation of the 
fine structure constant in accordance with observational data.

\section{Conclusion}

In the present paper, 
we have studied a domain wall solution in $F(R)$ gravity. 
We have reconstructed a static domain wall solution 
in a scalar field theory. 
We have also reconstructed an explicit $F(R)$ gravity model in which 
a static domain wall solution can be realized. 
Furthermore, 
we have shown that there could exists an effective (gravitational) 
domain wall in the framework of $F(R)$ gravity. 
Moreover, it has been illustrated that a logarithmic 
non-minimal gravitational coupling of the electromagnetic theory 
in $F(R)$ gravity may produce time-variation of the fine structure constant 
which may increase as the curvature decreases. 
In addition, 
we have described cosmological consequences of the coupling of the 
electromagnetic field to not only a scalar field but also 
the scalar curvature 
and remarked the relation between variation of the fine structure constant and 
the breaking of the conformal invariance of the electromagnetic field. 

The reconstruction technique was applied here to inducing of domain wall 
solution in modified gravity (cf. the case of black hole reconstruction in 
Ref.~\cite{Nojiri:2006jy}). It is clear that similar 
methods may be applied to generation of other solutions in modified 
gravity, like topological defects, cosmic strings, etc. This question will 
be discussed elsewhere.

\section*{Acknowledgments}

The work is supported in part
by Global COE Program
of Nagoya University (G07) provided by the Ministry of Education, Culture,
Sports, Science \& Technology
(S.N.); 
the JSPS Grant-in-Aid for Scientific Research (S) \# 22224003 and (C) 
\# 23540296 (S.N.); 
%
%
and
MEC (Spain) project FIS2006-02842 and AGAUR (Catalonia) 2009SGR-994
(S.D.O.).




\begin{thebibliography}{99}

\bibitem{SN1}
%
 S.~Perlmutter {\it et al.}  [SNCP Collaboration],
 Astrophys.\ J.\  {\bf 517}, 565 (1999) 
 [arXiv:astro-ph/9812133]; \\ 
%
  A.~G.~Riess {\it et al.}  [Supernova Search Team Collaboration],
  Astron.\ J.\  {\bf 116}, 1009 (1998)
  [arXiv:astro-ph/9805201].
%

\bibitem{WMAP}
%
 D.~N.~Spergel {\it et al.}  [WMAP Collaboration],
 Astrophys.\ J.\ Suppl.\  {\bf 148}, 175 (2003) 
 [arXiv:astro-ph/0302209]; \\ 
%
 D.~N.~Spergel {\it et al.}  [WMAP Collaboration],
 Astrophys.\ J.\ Suppl.\  {\bf 170}, 377 (2007) 
 [arXiv:astro-ph/0603449]; \\
%
 E.~Komatsu {\it et al.}  [WMAP Collaboration],
 Astrophys.\ J.\ Suppl.\  {\bf 180}, 330 (2009) 
 [arXiv:0803.0547 [astro-ph]]. 
%

\bibitem{Komatsu:2010fb}
  E.~Komatsu {\it et al.}  [WMAP Collaboration],
  Astrophys.\ J.\ Suppl.\  {\bf 192}, 18 (2011)
  [arXiv:1001.4538 [astro-ph.CO]].

\bibitem{LSS}
%
  M.~Tegmark {\it et al.}  [SDSS Collaboration],
  Phys.\ Rev.\  D {\bf 69}, 103501 (2004)
  [arXiv:astro-ph/0310723]; \\
%
  U.~Seljak {\it et al.}  [SDSS Collaboration],
  Phys.\ Rev.\  D {\bf 71}, 103515 (2005)
  [arXiv:astro-ph/0407372].
%

\bibitem{Eisenstein:2005su}
  D.~J.~Eisenstein {\it et al.}  [SDSS Collaboration],
  Astrophys.\ J.\  {\bf 633}, 560 (2005)
  [arXiv:astro-ph/0501171].

\bibitem{Jain:2003tba}
  B.~Jain and A.~Taylor,
  Phys.\ Rev.\ Lett.\  {\bf 91}, 141302 (2003)
  [arXiv:astro-ph/0306046].

\bibitem{Li:2011sd}
  M.~Li, X.~D.~Li, S.~Wang and Y.~Wang,
  Commun.\ Theor.\ Phys.\  {\bf 56}, 525 (2011)
  [arXiv:1103.5870 [astro-ph.CO]].

\bibitem{Review-Nojiri-Odintsov}
%
  S.~Nojiri and S.~D.~Odintsov,
  Phys.\ Rept.\  {\bf 505}, 59 (2011)
  [arXiv:1011.0544 [gr-qc]]; \\
%
 S.~Nojiri and S.~D.~Odintsov,
 eConf {\bf C0602061}, 06 (2006)
 [Int.\ J.\ Geom.\ Meth.\ Mod.\ Phys.\  {\bf 4}, 115 (2007)]
 [arXiv:hep-th/0601213].

\bibitem{Book-Capozziello-Faraoni}
S.~Capozziello and V.~Faraoni,
\textit{Beyond Einstein Gravity}
(Springer, 2010). 

\bibitem{Clifton:2011jh}
  T.~Clifton, P.~G.~Ferreira, A.~Padilla and C.~Skordis,
  arXiv:1106.2476 [astro-ph.CO].

\bibitem{Time-variation}
%
  J.~K.~Webb, V.~V.~Flambaum, C.~W.~Churchill, M.~J.~Drinkwater and J.~D.~Barrow,
  Phys.\ Rev.\ Lett.\  {\bf 82}, 884 (1999)
  [arXiv:astro-ph/9803165]; \\
%
  J.~K.~Webb {\it et al.},
  Phys.\ Rev.\ Lett.\  {\bf 87}, 091301 (2001)
  [arXiv:astro-ph/0012539]. 
%

%
\bibitem{Murphy:2003hw}
  M.~T.~Murphy, J.~K.~Webb and V.~V.~Flambaum,
  Mon.\ Not.\ Roy.\ Astron.\ Soc.\  {\bf 345}, 609 (2003)
  [arXiv:astro-ph/0306483].
%

\bibitem{Webb:2010hc}
  J.~K.~Webb, J.~A.~King, M.~T.~Murphy, V.~V.~Flambaum, R.~F.~Carswell and M.~B.~Bainbridge,
  arXiv:1008.3907 [astro-ph.CO].

\bibitem{Olive:2010vh}
  K.~A.~Olive, M.~Peloso and J.~P.~Uzan,
  Phys.\ Rev.\  D {\bf 83}, 043509 (2011)
  [arXiv:1011.1504 [astro-ph.CO]].

\bibitem{Chiba:2011en}
  T.~Chiba and M.~Yamaguchi,
  JCAP {\bf 1103}, 044 (2011)
  [arXiv:1102.0105 [astro-ph.CO]].

\bibitem{Cho:1998jk}
  I.~Cho and A.~Vilenkin,
  Phys.\ Rev.\  D {\bf 59}, 021701 (1998)
  [arXiv:hep-th/9808090].


\bibitem{M-B-C}
%
  J.~D.~Barrow and D.~F.~Mota,
  Class.\ Quant.\ Grav.\  {\bf 20}, 2045 (2003)
  [arXiv:gr-qc/0212032]; \\
%
  D.~F.~Mota and J.~D.~Barrow,
  Mon.\ Not.\ Roy.\ Astron.\ Soc.\  {\bf 349}, 291 (2004)
  [arXiv:astro-ph/0309273]; \\
%
  D.~F.~Mota and J.~D.~Barrow,
  Phys.\ Lett.\  B {\bf 581}, 141 (2004)
  [arXiv:astro-ph/0306047]; \\
%
  T.~Clifton, D.~F.~Mota and J.~D.~Barrow,
  Mon.\ Not.\ Roy.\ Astron.\ Soc.\  {\bf 358}, 601 (2005)
  [arXiv:gr-qc/0406001].
%

\bibitem{Chameleon-mechanism}
%
  J.~Khoury and A.~Weltman,
  Phys.\ Rev.\ Lett.\  {\bf 93}, 171104 (2004)
  [arXiv:astro-ph/0309300]; \\
%
  J.~Khoury and A.~Weltman,
  Phys.\ Rev.\  D {\bf 69}, 044026 (2004) 
  [arXiv:astro-ph/0309411].
%

\bibitem{Capozziello:2005tf}
  S.~Capozziello, S.~Nojiri and S.~D.~Odintsov,
  Phys.\ Lett.\  B {\bf 632}, 597 (2006)
  [arXiv:hep-th/0507182].

%
\bibitem{Reconstruction-F(R)-N-O}
%
  S.~Nojiri and S.~D.~Odintsov,
  Phys.\ Rev.\  D {\bf 74}, 086005 (2006)
  [arXiv:hep-th/0608008]; \\ 
%
  S.~Nojiri and S.~D.~Odintsov,
  J.\ Phys.\ Conf.\ Ser.\  {\bf 66}, 012005 (2007)
  [arXiv:hep-th/0611071]. 
%

\bibitem{F-M}
%
Y.~Fujii and K.~Maeda,
\textit{The Scalar-Tensor Theory of Gravitation}
(Cambridge University Press, Cambridge, United Kingdom, 2003); \\ 
%
  K.~I.~Maeda,
  Phys.\ Rev.\  D {\bf 39}, 3159 (1989).
%

\bibitem{Kolb and Turner}
E.~W.~Kolb and M.~S.~Turner,
\textit{The Early Universe}
(Addison-Wesley, Redwood City, California, 1990).

\bibitem{Freedman:2000cf}
 W.~L.~Freedman {\it et al.}  [HST Collaboration],
 Astrophys.\ J.\  {\bf 553} (2001) 47 
 [arXiv:astro-ph/0012376].

\bibitem{Exponential-Gravity}
%
  G.~Cognola, E.~Elizalde, S.~Nojiri, S.~D.~Odintsov, L.~Sebastiani and 
S.~Zerbini,
  Phys.\ Rev.\  D {\bf 77}, 046009 (2008)
  [arXiv:0712.4017 [hep-th]]; \\
%
  E.~V.~Linder,
  Phys.\ Rev.\  D {\bf 80}, 123528 (2009)
  [arXiv:0905.2962 [astro-ph.CO]]; \\
%
  K.~Bamba, C.~Q.~Geng and C.~C.~Lee,
  JCAP {\bf 1008}, 021 (2010)
  [arXiv:1005.4574 [astro-ph.CO]]; \\
%
  E.~Elizalde, S.~Nojiri, S.~D.~Odintsov, L.~Sebastiani and S.~Zerbini,
  Phys.\ Rev.\  D {\bf 83}, 086006 (2011)
  [arXiv:1012.2280 [hep-th]].
%

\bibitem{Drummond:1979pp}
  I.~T.~Drummond and S.~J.~Hathrell,
  Phys.\ Rev.\  D {\bf 22}, 343 (1980).

%

%
\bibitem{Bamba:2008ja}
  K.~Bamba and S.~D.~Odintsov,
  JCAP {\bf 0804}, 024 (2008)
  [arXiv:0801.0954 [astro-ph]]. 
%

%
\bibitem{Bamba:2008ut}
  K.~Bamba, S.~Nojiri and S.~D.~Odintsov,
  JCAP {\bf 0810}, 045 (2008)
  [arXiv:0807.2575 [hep-th]].
%

%
\bibitem{Bamba:2008xa}
  K.~Bamba, S.~Nojiri and S.~D.~Odintsov,
  Phys.\ Rev.\  D {\bf 77}, 123532 (2008)
  [arXiv:0803.3384 [hep-th]]. 
%

\bibitem{Elizalde:1996am}
  E.~Elizalde, S.~D.~Odintsov and A.~Romeo,
  Phys.\ Rev.\  D {\bf 54}, 4152 (1996)
  [arXiv:hep-th/9607189].



\bibitem{Turner:1987bw}
  M.~S.~Turner and L.~M.~Widrow,
  Phys.\ Rev.\  D {\bf 37}, 2743 (1988).

\bibitem{Scalar-EM-Magnetic-fields}
%
  B.~Ratra,
  Astrophys.\ J.\  {\bf 391}, L1 (1992); \\
%
  W.~D.~Garretson, G.~B.~Field and S.~M.~Carroll,
  Phys.\ Rev.\  D {\bf 46}, 5346 (1992)
  [arXiv:hep-ph/9209238]; \\
%
  D.~Lemoine and M.~Lemoine,
  Phys.\ Rev.\  D {\bf 52}, 1955 (1995); \\
%
  M.~Gasperini, M.~Giovannini and G.~Veneziano,
  Phys.\ Rev.\ Lett.\  {\bf 75}, 3796 (1995)
  [arXiv:hep-th/9504083]; \\
%
  G.~B.~Field and S.~M.~Carroll,
  Phys.\ Rev.\  D {\bf 62}, 103008 (2000)
  [arXiv:astro-ph/9811206]; \\
%
  M.~Giovannini,
  Phys.\ Rev.\  D {\bf 64}, 061301 (2001)
  [arXiv:astro-ph/0104290]; \\
%
  M.~Giovannini,
  arXiv:astro-ph/0212346; \\
%
%
  K.~Bamba and J.~Yokoyama,
  Phys.\ Rev.\  D {\bf 69}, 043507 (2004)
  [arXiv:astro-ph/0310824]; \\
%
  K.~Bamba and J.~Yokoyama,
  Phys.\ Rev.\  D {\bf 70}, 083508 (2004)
  [arXiv:hep-ph/0409237]; \\
%
  M.~Giovannini,
  Phys.\ Lett.\  B {\bf 659}, 661 (2008)
  [arXiv:0711.3273 [astro-ph]]; \\
%
  J.~Martin and J.~Yokoyama,
  JCAP {\bf 0801}, 025 (2008)
  [arXiv:0711.4307 [astro-ph]]; \\
%
  K.~Bamba, N.~Ohta and S.~Tsujikawa,
  Phys.\ Rev.\  D {\bf 78}, 043524 (2008)
  [arXiv:0805.3862 [astro-ph]]; \\
%
  K.~Bamba, C.~Q.~Geng and S.~H.~Ho,
  JCAP {\bf 0811}, 013 (2008)
  [arXiv:0806.1856 [astro-ph]].
%

\bibitem{Magnetic-fields-Bamba-Sasaki}
%
  K.~Bamba and M.~Sasaki,
  JCAP {\bf 0702}, 030 (2007)
  [arXiv:astro-ph/0611701]; \\
%
  K.~Bamba,
  JCAP {\bf 0710}, 015 (2007)
  [arXiv:0710.1906 [astro-ph]].
%

\bibitem{Conformally-flat-FLRWspacetime}
%
  F.~D.~Mazzitelli and F.~M.~Spedalieri,
  Phys.\ Rev.\  D {\bf 52}, 6694 (1995) 
  [arXiv:astro-ph/9505140]; \\
%
  G.~Lambiase and A.~R.~Prasanna,
  Phys.\ Rev.\  D {\bf 70}, 063502 (2004) 
  [arXiv:gr-qc/0407071].
%

\bibitem{Barrow:2011ic}
  J.~D.~Barrow and C.~G.~Tsagas,
  Mon.\ Not.\ Roy.\ Astron.\ Soc.\  {\bf 414}, 512 (2011)
  [arXiv:1101.2390 [astro-ph.CO]].

\bibitem{Kandus:2010nw}
  A.~Kandus, K.~E.~Kunze and C.~G.~Tsagas,
  Phys.\ Rept.\  {\bf 505}, 1 (2011)
  [arXiv:1007.3891 [astro-ph.CO]].


\bibitem{Nojiri:2006jy}
  S.~Nojiri, S.~D.~Odintsov and H.~Stefancic,
  Phys.\ Rev.\  D {\bf 74}, 086009 (2006)
  [arXiv:hep-th/0608168].

\end{thebibliography}
\end{document}